\documentclass[twocolumn]{aastex62}
\pdfoutput=1
 
\usepackage{verbatim}
\usepackage[usenames, dvipsnames]{}

\newcommand{\HARPS}{HARPS}
\newcommand{\Kepler}{Kepler}
\newcommand{\TESS}{TESS}

\newcommand{\Mearth}{\ensuremath{M_\earth}}
\newcommand{\Rearth}{\ensuremath{R_\earth}}

\newcommand{\bjdtdb}{\ensuremath{\rm {BJD_{TDB}}}}

\newcommand{\fave}{\langle F \rangle}
\newcommand{\feh}{\ensuremath{\left[{\rm Fe}/{\rm H}\right]}}
\newcommand{\fluxcgs}{\ensuremath{10^9\,erg\,s^{-1}\,cm^{-2}}}
\newcommand{\gaia}{\emph{Gaia}}
\newcommand{\gcmc}{\ensuremath{\rm g\,cm^{-3}}}

\newcommand{\logg}{\ensuremath{\log g}}

\newcommand{\lsun}{\ensuremath{L_\sun}}

\newcommand{\mearth}{\ensuremath{M_\earth}}
\newcommand{\me}{\ensuremath{\,M_{\rm E}}}

\newcommand{\msun}{\ensuremath{M_\sun}}
\newcommand{\ms}{\ensuremath{\rm m\,s^{-1}}}

\newcommand{\rearth}{\ensuremath{R_\earth}}
\newcommand{\re}{\ensuremath{\,R_{\rm E}}}

\newcommand{\rhopl}{\ensuremath{\rho_{p}}}

\newcommand{\rsun}{\ensuremath{R_\sun}}

\newcommand{\target}{TOI-824}

\newcommand{\teff}{\ensuremath{T_{\rm eff}}}

\newcommand{\thisplanetb}{TOI-824~b}
\newcommand{\thisstar}{TOI-824}

\newcommand{\mplanetunc}{18.47 $\pm$ 1.84 \mearth}

\newcommand{\rplanetunc}{2.93 $\pm$ 0.20 \rearth}

\newcommand{\rhoplanetunc}{4.03$^{+0.98}_{-0.78}$ \gcmc}

\shorttitle{A hot Neptune around TOI-824}
\shortauthors{Burt et al. 2020}

\begin{document}

\title{TOI-824~b: A New Planet on the Lower Edge of the Hot Neptune Desert}

\author[0000-0002-0040-6815]{Jennifer~A.~Burt}
\altaffiliation{Juan Carlos Torres Fellow}
\affiliation{Jet Propulsion Laboratory, California Institute of Technology, 4800 Oak Grove Drive, Pasadena, CA 91109, USA}
\affiliation{Department of Physics and Kavli Institute for Astrophysics and Space Research, Massachusetts Institute of Technology, Cambridge, MA 02139, USA}

\author[0000-0002-5254-2499]{Louise~D.~Nielsen}
\affiliation{Observatoire de l'Universit\'e de Gen\`eve, 51 chemin des Maillettes, 1290 Versoix, Switzerland}

\author[0000-0002-8964-8377]{Samuel~N.~Quinn}
\affiliation{Center for Astrophysics ${\rm \mid}$ Harvard {\rm \&} Smithsonian, 60 Garden Street, Cambridge, MA 02138, USA}

\author[0000-0003-2008-1488]{Eric~E.~Mamajek}
\affiliation{Jet Propulsion Laboratory, California Institute of Technology, 4800 Oak Grove Drive, Pasadena, CA 91109, USA}

\author[0000-0003-0593-1560]{Elisabeth~C.~Matthews}
\affiliation{Department of Physics and Kavli Institute for Astrophysics and Space Research, Massachusetts Institute of Technology, Cambridge, MA 02139, USA}

\author[0000-0002-4891-3517]{George Zhou}
\affiliation{Center for Astrophysics ${\rm \mid}$ Harvard {\rm \&} Smithsonian, 60 Garden Street, Cambridge, MA 02138, USA}

\author[0000-0002-7990-9596]{Julia~V.~Seidel}
\affiliation{Observatoire de l'Universit\'e de Gen\`eve, 51 chemin des Maillettes, 1290 Versoix, Switzerland}

\author{Chelsea~ X.~Huang}
\altaffiliation{Juan Carlos Torres Fellow}
\affiliation{Department of Physics and Kavli Institute for Astrophysics and Space Research, Massachusetts Institute of Technology, Cambridge, MA 02139, USA}

\author{Eric~Lopez}
\affiliation{NASA Goddard Space Flight Center, 8800 Greenbelt Road, Greenbelt, MD 20771, USA}

\author[0000-0001-9743-5649]{Maritza Soto}
\affiliation{School of Physics and Astronomy, Queen Mary University London, 327 Mile End Road, London E1 4NS, United Kingdom}

\author{Jon Otegi}
\affiliation{Institute for Computational Science, University of Zurich, Winterthurerstr. 190, CH-8057 Zurich, Switzerland}
\affiliation{Observatoire de l'Universit\'e de Gen\`eve, 51 chemin des Maillettes, 1290 Versoix, Switzerland}

\author[0000-0002-3481-9052]{Keivan~G.~Stassun}
\affiliation{Department of Physics and Astronomy, Vanderbilt University, Nashville, TN 37235, USA}
\affiliation{Department of Physics, Fisk University, Nashville, TN 37208, USA}

\author{Laura~Kreidberg}
\altaffiliation{Clay Fellow}
\affiliation{Center for Astrophysics ${\rm \mid}$ Harvard {\rm \&} Smithsonian, 60 Garden Street, Cambridge, MA 02138, USA}

\author[0000-0001-6588-9574]{Karen~A.~Collins}
\affiliation{Center for Astrophysics ${\rm \mid}$ Harvard {\rm \&} Smithsonian, 60 Garden Street, Cambridge, MA 02138, USA}

\author[0000-0003-3773-5142]{Jason~D.~Eastman}
\affiliation{Center for Astrophysics ${\rm \mid}$ Harvard {\rm \&} Smithsonian, 60 Garden Street, Cambridge, MA 02138, USA}

\author[0000-0001-8812-0565]{Joseph~E.~Rodriguez}
\affiliation{Harvard--Smithsonian Center for Astrophysics, Harvard University, Cambridge, MA 02138, USA}

\author[0000-0001-7246-5438]{Andrew~Vanderburg}
\affiliation{Department of Astronomy, University of Wisconsin-Madison, Madison, WI 53706, USA}

\author[0000-0003-1312-9391]{Samuel~P.~Halverson}
\affiliation{Jet Propulsion Laboratory, California Institute of Technology, 4800 Oak Grove Drive, Pasadena, CA 91109, USA}

\author{Johanna~K.~Teske}
\altaffiliation{NASA Hubble Fellow}
\affiliation{The Observatories of the Carnegie Institution for Science, 813 Santa Barbara Street, Pasadena, CA 91101, USA}

\author[0000-0002-6937-9034]{Sharon~X.~Wang}
\affiliation{The Observatories of the Carnegie Institution for Science, 813 Santa Barbara Street, Pasadena, CA 91101, USA}

\author[0000-0003-1305-3761]{R.~Paul~Butler}
\affiliation{Earth \& Planets Laboratory, Carnegie Institution for Science, 5241 Broad Branch Road, NW, Washington, DC 20015, USA}

\author{Fran\c{c}ois~Bouchy}
\affiliation{Observatoire de l'Universit\'e de Gen\`eve, 51 chemin des Maillettes, 1290 Versoix, Switzerland}

\author[0000-0002-9332-2011]{Xavier Dumusque}
\affiliation{Observatoire de l'Universit\'e de Gen\`eve, 51 chemin des Maillettes, 1290 Versoix, Switzerland}

\author{Damien Segransen}
\affiliation{Observatoire de l'Universit\'e de Gen\`eve, 51 chemin des Maillettes, 1290 Versoix, Switzerland}

\author{Stephen~A.~Shectman}
\affiliation{The Observatories of the Carnegie Institution for Science, 813 Santa Barbara Street, Pasadena, CA 91101, USA}

\author[0000-0002-5226-787X]{Jeffrey~D.~Crane}
\affiliation{The Observatories of the Carnegie Institution for Science, 813 Santa Barbara Street, Pasadena, CA 91101, USA}

\author[0000-0001-6039-0555]{Fabo Feng}
\affiliation{Earth \& Planets Laboratory, Carnegie Institution for Science, 5241 Broad Branch Road, NW, Washington, DC 20015, USA}

\author[0000-0001-7516-8308]{Benjamin~T.~Montet}
\affiliation{School of Physics, The University of New South Wales, Sydney, NSW 2052, Australia}

\author[0000-0002-9464-8101]{Adina~D.~Feinstein}
\altaffiliation{NSF Graduate Research Fellow}
\affiliation{The Department of Astronomy and Astrophysics, The University of Chicago, 5640 S. Ellis Ave, Chicago, IL 60637, USA}

\author{Yuri Beletski}
\affiliation{Las Campanas Observatory, Vallenar, Atacama, Chile}

\author[0000-0001-8045-1765]{Erin Flowers}
\altaffiliation{NSF Graduate Research Fellow}
\affiliation{Department of Astrophysical Sciences, Princeton University, 4 Ivy Lane, Princeton, NJ 08544, USA}

\author{Maximilian~N.~G{\"u}nther}
\altaffiliation{Juan Carlos Torres Fellow}
\affiliation{Department of Physics and Kavli Institute for Astrophysics and Space Research, Massachusetts Institute of Technology, Cambridge, MA 02139, USA}

\author[0000-0002-6939-9211]{Tansu~Daylan}
\altaffiliation{Kavli Fellow}
\affiliation{Department of Physics and Kavli Institute for Astrophysics and Space Research, Massachusetts Institute of Technology, Cambridge, MA 02139, USA}

\author[0000-0003-2781-3207]{Kevin~I.~Collins}
\affiliation{George Mason University, 4400 University Drive, Fairfax, VA, 22030 USA}

\author[0000-0003-2239-0567]{Dennis~M.~Conti}
\affiliation{American Association of Variable Star Observers, 49 Bay State Road, Cambridge, MA 02138, USA}

\author[0000-0002-4503-9705]{Tianjun~Gan}
\affiliation{Department of Astronomy and Tsinghua Centre for Astrophysics, Tsinghua University, Beijing 100084, China}

\author[0000-0002-4625-7333]{Eric~L.~N.~Jensen}
\affiliation{Dept.\ of Physics \& Astronomy, Swarthmore College, Swarthmore PA 19081, USA}

\author[0000-0003-0497-2651]{John~F.~Kielkopf}
\affiliation{Department of Physics and Astronomy, University of Louisville, Louisville, KY 40292, USA}

\author[0000-0001-5603-6895]{Thiam-Guan~Tan}
\affiliation{Perth Exoplanet Survey Telescope, Perth, Western Australia, Australia}

\author[0000-0001-5555-2652]{Ravit~Helled}
\affiliation{Institute for Computational Science, University of Zurich, Winterthurerstr. 190, CH-8057 Zurich, Switzerland}

\author[0000-0001-6110-4610]{Caroline~Dorn}
\affiliation{Institute for Computational Science, University of Zurich, Winterthurerstr. 190, CH-8057 Zurich, Switzerland}

\author[0000-0003-1231-2389]{Jonas~Haldemann}
\affiliation{Department of Space Research \& Planetary Sciences, University of Bern, Gesellschaftsstrasse 6, CH-3012 Bern, Switzerland}

\author{Jack~J.~Lissauer}
\affiliation{Space Science \& Astrobiology Division, MS 245-3, NASA Ames Research Center, Moffett Field, CA 94035, USA}

\author[0000-0003-2058-6662]{George~R.~Ricker}
\affiliation{Department of Physics and Kavli Institute for Astrophysics and Space Research, Massachusetts Institute of Technology, Cambridge, MA 02139, USA}

\author[0000-0001-6763-6562]{Roland~Vanderspek}
\affiliation{Department of Physics and Kavli Institute for Astrophysics and Space Research, Massachusetts Institute of Technology, Cambridge, MA 02139, USA}

\author[0000-0001-9911-7388]{David~W.~Latham}
\affiliation{Center for Astrophysics ${\rm \mid}$ Harvard {\rm \&} Smithsonian, 60 Garden Street, Cambridge, MA 02138, USA}

\author{S.~Seager}
\affiliation{Department of Physics and Kavli Institute for Astrophysics and Space Research, Massachusetts Institute of Technology, Cambridge, MA 02139, USA}
\affiliation{Department of Earth, Atmospheric and Planetary Sciences, Massachusetts Institute of Technology, Cambridge, MA 02139, USA}
\affiliation{Department of Aeronautics and Astronautics, MIT, 77 Massachusetts Avenue, Cambridge, MA 02139, USA}

\author{Joshua~N.~Winn}
\affiliation{Department of Astrophysical Sciences, Princeton University, 4 Ivy Lane, Princeton, NJ 08544, USA}

\author[0000-0002-4715-9460]{Jon~M.~Jenkins}
\affiliation{NASA Ames Research Center, Moffett Field, CA, 94035, USA}

\author[0000-0002-6778-7552]{Joseph~D.~Twicken}
\affiliation{NASA Ames Research Center, Moffett Field, CA, 94035, USA}
\affiliation{SETI Institute, Mountain View, CA 94043, USA}

\author[0000-0002-6148-7903]{Jeffrey~C.~Smith}
\affiliation{NASA Ames Research Center, Moffett Field, CA, 94035, USA}
\affiliation{SETI Institute, Mountain View, CA 94043, USA}

\author[0000-0002-1949-4720]{Peter~Tenenbaum}
\affiliation{NASA Ames Research Center, Moffett Field, CA, 94035, USA}
\affiliation{SETI Institute, Mountain View, CA 94043, USA}

\author{Scott~Cartwright}
\affiliation{Proto-Logic LLC, 1718 Euclid Street NW, Washington, DC 20009, USA}

\author[0000-0001-7139-2724]{Thomas~Barclay}
\affiliation{NASA Goddard Space Flight Center, 8800 Greenbelt Road, Greenbelt, MD 20771, USA}
\affiliation{University of Maryland, Baltimore County, 1000 Hilltop Circle, Baltimore, MD 21250, USA}

\author[0000-0002-3827-8417]{Joshua~Pepper}
\affiliation{Department of Physics, Lehigh University, 16 Memorial Drive East, Bethlehem, PA 18015, USA}

\author{Gilbert~Esquerdo}
\affiliation{Center for Astrophysics ${\rm \mid}$ Harvard {\rm \&} Smithsonian, 60 Garden Street, Cambridge, MA 02138, USA}

\author[0000-0003-0241-2757]{William~Fong}
\affiliation{Department of Physics and Kavli Institute for Astrophysics and Space Research, Massachusetts Institute of Technology, Cambridge, MA 02139, USA}

\begin{abstract}
We report the detection of a transiting hot Neptune exoplanet orbiting \object{TOI-824} (\object{SCR J1448-5735}), a nearby ($d$ = 64 pc) K4V star, using data from the \textit{Transiting Exoplanet Survey Satellite} (\TESS). The newly discovered planet has a radius, $R_p$  = \rplanetunc, and an orbital period of 1.393 days. Radial velocity measurements using the Planet Finder Spectrograph (PFS) and the High Accuracy Radial velocity Planet Searcher (HARPS) spectrograph confirm the existence of the planet and we estimate its mass to be \mplanetunc. The planet's mean density is \rhopl\, = \rhoplanetunc, making it more than twice as dense as Neptune. \object{TOI-824 b}'s high equilibrium temperature makes the planet likely to have a cloud free atmosphere, and thus an excellent candidate for follow up atmospheric studies. The detectability of TOI-824 b's atmosphere from both ground and space is promising and could lead to the detailed characterization of the most irradiated, small planet at the edge of the hot Neptune desert that has retained its atmosphere to date. 
\end{abstract}

\section{Introduction}

The {\it Transiting Exoplanet Survey Satellite} \citep[\TESS,][]{ricker} spent the first year of its mission searching for planets orbiting cool, nearby stars in the Ecliptic Southern Hemisphere. To date, the \TESS\ mission has detected over 1000 planet candidates and has significantly expanded the number of small planets detected around cool stars \citep[see, e.g.,][]{Gunther2019, Kostov2019, Dragomir2019}. 

There are a number of striking planet populations that emerge when studying the period and radius measurements that \TESS\, and its predecessor \emph{Kepler}, have compiled for the 2000+ confirmed, transiting exoplanets they detected. One of the most surprising is the huge population of planets between the size of Earth and Neptune that orbit stars of all stellar types \citep{Coughlin2016}, a population that is missing from our own solar system. Equally interesting and enigmatic are the hot Jupiters that orbit their stars with periods thousands of times shorter than our own Jupiter \citep[e.g.][]{MayorQueloz1995} and the small, tightly packed, transiting planets around FGKM dwarfs that often exist in low mutual inclination systems just wide of orbital resonance chains \citep{Lissauer2011, Burke2014}. At the same time, this sizeable data set reveals the lack of planets within certain regions of parameter space. One notable example is the existence of the ``hot Neptune desert", or the lack of planets the size and mass of Neptune on periods shorter than 4 days that is seen in both transit and Doppler detections \citep{SzaboKiss2011, Mazeh2016}. This desert cannot be a result of observational bias as \Kepler, \TESS\, and various radial velocity (RV) surveys have detected a plethora of planets with similar masses and radii to Neptune on much longer period orbits, which have lower transit detection probabilities and smaller RV semi-amplitudes \citep{Lecavelier2007, Beauge2013}. Thus these hot Neptunes must be intrinsically rare, and indeed analysis of the \Kepler\ DR25 planet candidates suggests that \textless 1\%\ of FGK stars host planets that have radii in the range of 2-6\rearth\ on orbits shorter than 4 days \citep{Hsu2019}. 

The origin of these hot Neptunes remains unclear as their radii lie between the terrestrial planets that are thought to form primarily in-situ \citep{MatsumotoKokubo2017} and the giant planets that astronomers have traditionally believed must form out past the snow line before migrating inwards \citep{Nelson2017}. 

Two characteristics that have been noted are that hot Neptunes, much like hot Jupiters, are more likely to be found around metal-rich stars and are more likely to be found in systems where they are the only transiting planet \citep{Dong2018, Petigura2018}. Determining whether or not these similarities suggest a common origin method or migration history between the two hot planet populations requires the detection and confirmation of enough hot Neptunes to identify which planetary, atmospheric, and orbital characteristics are common across hot Neptunes and which vary from planet to planet. Thanks to the inherent rarity of these planets in the Milky Way, the only way to compile such a set is to search a very large number of stars for evidence of hot Neptunes. Such a survey is impractical for RV instruments, where only one star can be observed at a time. Even \Kepler, which studied over 100,000 stars at a precision level that should easily detect such planets, discovered only a handful \citep{Hsu2019}. With its high level of photometric precision and an observing plan that will tile over 70\% of the night sky for at least 28 days during its primary and extended missions, \TESS\ is an ideal mission to detect hot Neptunes. Indeed, during its first year alone the mission has already added three verified hot Neptune planets to the exoplanet databases \citep[][and this work]{Esposito2019, Diaz2019} and identified over 100 additional hot Neptune planet candidates that await confirmation\footnote{https://tev.mit.edu/data/}.

Here we report the discovery of a hot Neptune orbiting \object{TOI-824}, a nearby ($d$ = 64 pc) K4 dwarf star located in the constellation of Circinus. This paper is organized as follows. In Section 2 we describe the variety of data sets used to characterize \object{TOI-824} and its planet. In Section 3 we detail the analysis of these data sets, culminating in the use of the EXOFASTv2 software package to determine the system's stellar, orbital, and planetary parameters. In Section 4 we detail how including ground based photometry in our fits revealed that the initial Science Processing Operations Center (SPOC) radius estimate for TOI-824 b was $\sim$11\% too large, and we outline suggestions for future \TESS\ follow up efforts focusing on stars in crowded regions of the sky. Finally, we conclude in Section 5 with a discussion of the planet's likely interior composition, the atmospheric retention capabilities of this and other hot Neptune planets, and TOI-824 b's and potential for future atmospheric characterization efforts.


\section{Data \label{sec:data}}

\subsection{Astrometry \& Photometry}

\object{TOI-824} (SCR J1448-5735, TYC 8688-915-1, TIC 193641523, 2MASS J14483982-5735175) is a $V$ = 11.15 \citep{Winters2011} K-type dwarf star located at 64.4$\pm$0.03 pc \citep[$\varpi$\, =\,15.6142\, $\pm$\, 0.0348 mas;][]{GaiaDR2}. 
The star has not drawn much previous attention, having been first pointed out as a high proper motion, potentially nearby star by \citet{Finch2007} (IDed as SCR J1448-5735).
As the star had a preliminary photometric distance estimate of only 18 pc by \citet{Finch2007}, \citet{Winters2011} measured Krons-Cousins $VRI$ photometry as part of the SuperCOSMOS RECONS survey, and provided an updated distance estimate of 62.4\,$\pm$\,9.7 pc, not far from the current Gaia DR2 estimate \citep{Bailer-Jones:2018}.
Astrometry and photometry for TOI-824 is summarized in Table \ref{tab:star}. 
The star's position, proper motion, parallax, and Gaia photometry are drawn from Gaia DR2. 
Optical photometry is adopted from 
Tycho-2 \citep[$B_T V_T$;][]{Hog2000},
APASS DR9 \citep[$BVr'i'$][]{Henden2015,Henden2016}, 
Gaia DR2 \citep{GaiaDR2}, 
while infrared photometry is adopted from 
2MASS \citep[$JHK_s$;][]{Cutri2003} and 
WISE \citep{Cutri2012}.  

\begin{table}[htbp]
    \caption{Stellar parameters for TOI-824.}\label{tab:star}
    \setlength{\tabcolsep}{3pt}
    \resizebox{\columnwidth}{!}{
    \begin{tabular}{lcc}
    \hline\hline
    Parameter     & Value & Source \\
     \hline \hline
    Designations & TIC 193641523  & \citet{StassunTIC2019}\\
                 & SCR J1448-5735 & \citet{Finch2007}\\
                 & 2MASS J14483982-5735175 & \citet{Cutri2003}\\
    R.A. (hh:mm:ss) & 14:48:39.71 & Gaia DR2\\
    Decl. (dd:mm:ss) & -57:35:19.92 & Gaia DR2\\
    $\mu$ R.A. (mas yr$^{-1}$) & -51.6035 $\pm$ 0.068 & Gaia DR2\\
    $\mu$ Decl. (mas yr$^{-1}$)& -152.392 $\pm$ 0.079 & Gaia DR2\\
    Parallax (mas) & 15.696 $\pm$ 0.049 & Gaia DR2*\\
    Distance (pc) & 63.71 $\pm$ 0.20 & Gaia DR2\\
    SpT & K4V & This work \\
    \hline
     $B$ & 12.263 $\pm$ 0.005 & APASS/DR9\\
     $V$ & 11.153 $\pm$ 0.008 & APASS/DR9\\
     $V$ & 11.15 $\pm$ 0.03: & \citet{Winters2011}\\
     $r'$ & 10.717 $\pm$ 0.025 & APASS/DR9\\
     $i'$ & 10.303 $\pm$ 0.053 & APASS/DR9\\
      TESS & 10.0732 $\pm$ 0.006 & TIC8\\
      $G$ & 10.762$^{+0.02}_{-0.001}$ & Gaia DR2 \\
      $G_{\rm{BP}}$ & 11.406$^{+0.02}_{-0.001}$ & Gaia DR2 \\
      $G_{\rm{RP}}$ & 10.021$^{+0.02}_{-0.001}$ & Gaia DR2 \\
      $J$  & 9.145 $\pm$ 0.018 & 2MASS\\
      $H$ & 8.557 $\pm$ 0.046 & 2MASS\\
      $K_{s}$ & 8.432 $\pm$ 0.038 & 2MASS\\
      W$_{1}$& 8.154 $\pm$ 0.083 & WISE \\
      W$_{2}$& 8.326 $\pm$ 0.021 & WISE \\
      W$_{3}$& 6.948 $\pm$ 0.017 & WISE \\
      W$_{4}$& 3.372 $\pm$ 0.020 & WISE \\
\hline
    $U$ (km s$^{-1}$) & -0.220 $\pm$ 0.152 & This work\\ 
    $V$ (km s$^{-1}$) & -45.803 $\pm$ 0.181 & This work\\
    $W$ (km s$^{-1}$) & -33.886 $\pm$ 0.140 & This work\\ 
\hline
\hline
    \multicolumn{3}{l}{*Correction of +82 $\mu$as applied to the Gaia parallax} \\ 
    \multicolumn{3}{l}{as per \citet{StassunTorres:2018}}\\
    \end{tabular}}
\end{table}

\subsection{\TESS\ Time Series Photometry}

\thisstar\ was observed by \TESS\ from UT 22 April 2019 through 20 May 2019 as part of the Sector 11 campaign and again from UT 21 May 2019 through 18 June 2019 as part of Sector 12. The star fell on CCD~3 of Camera~2 during Sector 11 and then on CCD~4 of Camera~2 during Sector 12.

\begin{figure*}
\includegraphics[width=\linewidth]{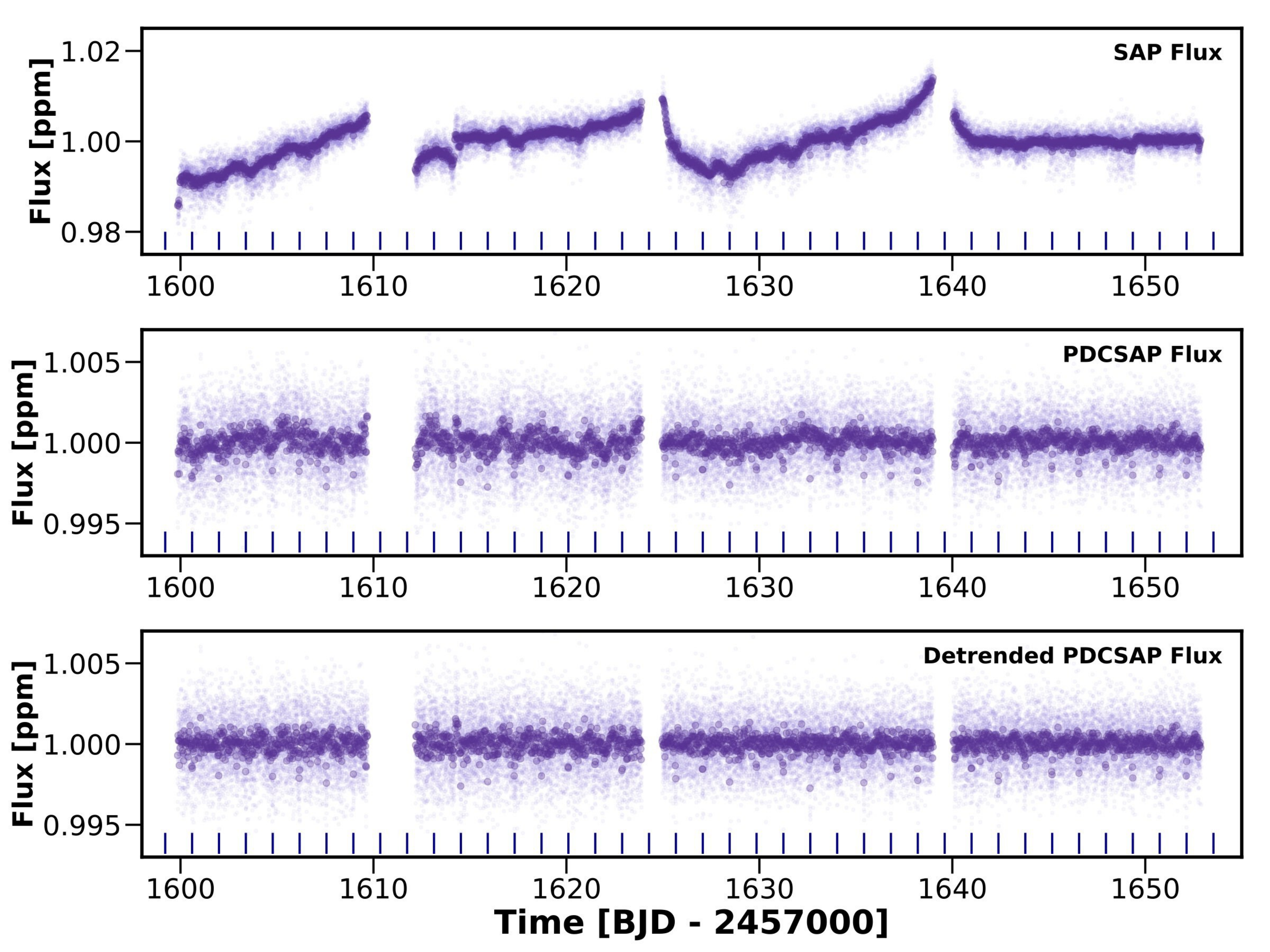}
\caption{SAP (\emph{top}), PDCSAP (\emph{middle}), and detrended PDCSAP (\emph{bottom}) \TESS\ light curves. The lighter points are the \TESS\ two-minute cadence flux measurements. The darker points are the same data binned over 30 minutes. The dark blue line segments at the bottom of each panel represent the locations of the transit events identified in the \TESS\ Data Validation Report based on the 2 minute cadence data.
\label{fig:lightcurve}}
\vspace{0.7cm}
\end{figure*}

The Science Processing Operations Center (SPOC) data \citep{Jenkins2016} for \thisstar\, available at the the Mikulski Archive for Space Telescopes (MAST) website\footnote{https://mast.stsci.edu}, includes both simple aperture photometry (SAP) flux measurements \citep{Twicken2010, Morris2017} and presearch data conditioned simple aperture photometry (PDCSAP) flux measurements \citep{Smith2012, Stumpe2012, Stumpe2014}. The instrumental variations present in the SAP flux are removed in the PDCSAP result. At the start of each orbit, thermal effects and strong scattered light impact the systematic error removal in PDC (see \TESS\ data release note DRN16 and DNR17). Before the fitting process described in Section \ref{sec:exofastv2}, we use the quality flags provided by SPOC to mask out unreliable segments of the time series. We then further detrend the \TESS\ data set by breaking it into the individual spacecraft orbits (two per sector) and fitting each with a low order spline to address residual trends in the light curve (Figure \ref{fig:lightcurve}). 

\subsection{Ground-based Time-Series Photometry}

We acquired ground-based time-series follow-up photometry of \thisstar\ during the times of transit predicted by the \TESS\ data. We used the {\tt TESS Transit Finder}, which is a customized version of the {\tt Tapir} software package \citep{Jensen:2013}, to schedule our transit observations.

We observed a full transit of \thisstar\ on UTC 2019 July 01 in $\rm R_c$ band from the Perth Exoplanet Survey Telescope (PEST) near Perth, Australia. The 0.3 m telescope is equipped with a $1530\times1020$ SBIG ST-8XME camera with an image scale of 1$\farcs$2 pixel$^{-1}$ resulting in a $31\arcmin\times21\arcmin$ field of view. A custom pipeline based on {\tt C-Munipack}\footnote{http://c-munipack.sourceforge.net} was used to calibrate the images and extract differential photometry for \thisstar\ and all other stars detected within $2\farcm5$ of \thisstar, using apertures with radius 6$\farcs$2. The images have typical stellar point spread functions (PSFs) with a FWHM of 3$\farcs$6.  

Four full transits of \thisplanetb\ were observed using the Las Cumbres Observatory Global Telescope (LCOGT) 1~m network \citep{Brown:2013} on UTC 2019 June 30, 2019 July 13, 2019 July 27, and 2019 August 05 in Pan-STARSS $z$, B, B, and Pan-STARSS $z$ bands, respectively. All observations were obtained by the LCOGT node at Cerro Tololo Inter-American Observatory, except the August 05 observations which were obtained by the LCOGT node at Siding Spring Observatory. The telescopes are equipped with $4096\times4096$ LCO SINISTRO cameras having an image scale of 0$\farcs$389 pixel$^{-1}$ resulting in a $26\arcmin \times 26\arcmin$ field of view. The images were calibrated by the standard LCOGT BANZAI pipeline and the photometric data were extracted using the {\tt AstroImageJ} ({\tt AIJ}) software package \citep{Collins:2017}. Circular apertures with radius 12 pixels (4$\farcs$7) were used to extract the differential photometry. The image sets have average stellar PSF FWHMs ranging from $1\farcs4$ on August 5th to $2\farcs4$ on July 14th.

\subsection{Spectroscopic Data}

\target\ was observed by the CHIRON spectrograph \citep{Tokovinin:2013} to determine whether its stellar parameters were well suited to precision RV follow up efforts. CHIRON is a high resolution spectrograph fed by an image slicer and a fiber bundle, located on the 1.5\,m SMARTS telescope at Cerro Tololo Inter-American Observatory (CTIO), Chile. CHIRON has a spectral resolving power of $R \simeq$ 80,000 over the wavelength region from 4100 to 8700\,\AA. Two spectra were obtained for \target\ on UT 5 and 6 July 2019. 

\subsection{Imaging Data}

We collected AO images with VLT/NaCo on UT 2 August 2019, using the Br$\gamma$ filter centered on 2.166$\mu$m \citep{Lenzen2003, Rousset2003}. We collected a total of 9 frames, each with an exposure time of 22 seconds, and dithered the telescope position between each frame. This allows a sky background to be constructed by median combining the dithered frames. A standard data reduction was carried out using a custom IDL code. This procedure included bad pixel removal, flat fielding and subtraction of the sky background. Frames were then aligned and co-added. 

\subsection{Time Series Radial Velocities}

\target\, was added as a target in two radial velocity (RV) \TESS\ follow up efforts using the Planet Finder Spectrograph (PFS) on the 6.5m Magellan Clay telescope \citep{Crane2006, Crane2008, Crane2010} at Las Campanas Observatory and the High-Accuracy Radial-velocity Planet Searcher \HARPS\ spectrograph \citep{Mayor2003} on the ESO~3.6m telescope at La Silla Observatory. 

PFS is a custom designed precision RV echelle spectrometer that, with the exception of the focus, has no moving parts. PFS is embedded in an insulated box where the temperature is maintained at 27$^{\circ}$C $\pm$ 0.01$^{\circ}$C. The wavelength range extends from 3900 to 6700\AA. In January 2018 the old PFS CCD (a 4K x 4K detector with 15$\mu$m pixels) was replaced with a next generation 10K x 10K detector with 9$\mu$m pixels, improving the sampling by 40\%. The peak resolution of PFS when using the 0.3" slit, as was done for all observations of TOI-824, is $R \simeq$ 130,000. An iodine cell placed in the converging beam of the telescope is used to provide a precise wavelength metric for velocity measurements \citep{MarcyButler1992}. The gaseous iodine blankets the region from 5000 to 6200\AA\ with a dense forest of sharp absorption lines.  The PFS iodine cell was scanned with the NIST FTS spectrometer \citep{Nave2017} at a resolution of 1 million. The raw data is reduced to 1D spectra with a custom built raw reduction package. Velocities were generated from an updated version of the iodine modeling package outlined in \citet{Butler1996} and the BJD time stamps were computed using the PEXO software package \citep{pexo}.

In contrast, the HARPS spectrograph makes use of multiple observing fibers, one of which is placed on the stellar target while the other is fed by a Fabry-Perot interferometer for a simultaneous wavelength reference. HARPS produces spectra from 3800 to 6900\AA, the entirety of which can be used to measure a star's radial velocity shift, and has a peak resolving power of $R \simeq$ 115,000 \citep{Pepe2002}. Once an observation is complete, a 2-D spectrum is optimally extracted from the resulting FITS file. The spectrum is cross-correlated with a numerical mask corresponding to the appropriate spectral type (F0, G2, K0, K5, or M4; we used the G2 which has undergone the most development during HARPS' observing span), and the resulting cross-correlation function (CCF) is fit with a Gaussian curve to produce a radial velocity measurement \citep{Baranne1996, Pepe2002} and calibrated to determine the RV photon-noise uncertainty $\sigma_{\rm{RV}}$.

A total of 24 PFS radial observations were obtained in July and August 2019, binned into 12 velocity measurements, with a mean internal uncertainty of 0.94~\ms. A total of eight HARPS observations, binned into five velocity measurements, were obtained in July 2019, with a mean internal uncertainty of 2.24~\ms (Table \ref{tab:RVs}).

\begin{table}[htbp]
\centering
\caption{Binned RV data of TOI-824}
\begin{tabular}{cccc}
\hline\hline
Date (BJD) & RV ($\ms$) & $\sigma_{RV}$ ($\ms$) & Instrument \\
\hline \vspace{2pt}
2458674.64056 & 4.73 & 1.91 & HARPS \\
2458675.67088 & 22.39 & 2.57 & HARPS \\
2458676.56075 & 4.42 & 0.94 & PFS \\
2458677.54783 & -11.54 & 0.92 & PFS \\
2458679.57467 & 15.06 & 0.83 & PFS \\
2458680.55934 & -3.35 & 0.79 & PFS \\
2458681.56315 & -4.86 & 1.07 & PFS \\
2458682.54917 & 12.28 & 0.84 & PFS \\
2458684.64071 & 6.30 & 2.12 & HARPS \\
2458685.53003 & 11.15 & 1.03 & PFS \\
2458689.59665 & 24.17 & 2.31 & HARPS \\
2458690.61645 & 18.20 & 2.30 & HARPS \\
2458703.54142 & 7.28 & 0.97 & PFS \\
2458705.51575 & -11.65 & 0.98 & PFS \\
2458708.49699 & -4.44 & 0.73 & PFS \\
2458712.49177 & -16.84 & 1.18 & PFS \\
2458714.48836 & 14.16 & 1.03 & PFS \\
\hline\hline
\end{tabular}
\label{tab:RVs}
\end{table}


\section{Analysis \label{sec:analysis}}

\subsection{Transit Detection \label{sec:transitdetection}}

Sixteen transits of \thisplanetb\ were detected in both the MIT Quick Look Pipeline (QLP), which searches for evidence of planet candidates in the \TESS\ 30 minute cadence Full Frame Images, and in the Science Processing Operations Center (SPOC) pipeline, which analyzes the 2-minute cadence data that \TESS\ obtains for pre-selected target stars \citep{Jenkins:2016}. Initial analyses of the \TESS\ QLP results found a transit signal with a period of 1.393 days, a depth of $\sim$1900 ppm, a duration of 1.3 hours, and the classic flat-bottomed trough shape that is characteristic of a planetary transit. These transit parameters translated to a planet with radius $R_p$ = 3.4\rearth\ when using the stellar parameters for \thisstar\ listed in the eighth installment of the \TESS\ Input Catalog \citep[TIC;][]{StassunTIC2019}. 

Although the measured transit depth and flat bottomed shape were positive indicators of a transit signal being planetary in nature, the \TESS\ vetting process is designed to guard against a variety of false positives that can mimic this combination. The primary sources of false positives in the \TESS\ data are eclipsing binaries, whether in the form of transiting stars on grazing orbits or as background blends which reduce the amplitude of foreground transit signals causing them to be deceptively small \citep[e.g.][]{Cameron2012}. Thus during the \TESS\ vetting process we carefully inspected the star's Data Validation Report \citep[DVR,][]{Twicken2018,Li2019}, which is based upon the SPOC two minute cadence data for \thisstar. The multi-sector DVR was found to show no evidence of secondary eclipses, inconsistencies in depth between the even and odd transits, nor correlations between aperture size and transit depth, any of which would be interpreted as a sign of the signal being caused by a nearby eclipsing binary. Additionally the DVR shows that the location of the transit source is consistent with the position of the target star. After passing these verification steps, the transit signal was assigned the identifier TOI-824.01 and was announced on the MIT \TESS\ data alerts website\footnote{http://tess.mit.edu/alerts} so that additional follow up efforts could be coordinated.

\subsection{Confirming the Source of the Transit Detection}

\TESS\ has large $\sim21\arcsec$ pixels and $\sim1\arcmin$ stellar FWHM resulting in photometric apertures that typically extend $\sim1\arcmin$ from the central target star location. The large apertures are often contaminated with many nearby stars bright enough to produce the \TESS\ detection. We used our higher spatial resolution follow-up time-series images to search for the location of the periodic flux deficits that caused the detection of \thisplanetb\ in the \TESS\ data. We checked the light curves of all 338 \gaia\ DR2 stars within $2\farcm5$ of \thisstar\ that are bright enough to have caused the \TESS\ detection and ruled out nearby EBs as the source. Furthermore, we detected five transits of \thisplanetb\ in our ground-based data using target star apertures with radii as small as $1\farcs2$ centered on \thisstar. The nearest \gaia\ DR2 or TICv8 star to \thisstar\ at the epoch of our follow-up observations is $7\farcs7$ West. Thus most of the flux from known nearby stars is excluded from even our larger 4$\farcs$7 target star apertures. We therefore confirm that the source of the flux deficit that is responsible for the TESS detection occurs within a $1\farcs2$ radius of \thisstar. Our multi-band light curves in B, $\rm R_c$, and Pan-STARRS $z$ bands show that the transits have depths that are consistent across optical wavelengths and with the deblended depth in the \TESS\ data. This rules out certain classes of bound or background EBs that could be blended in the small follow-up photometric apertures as potential sources of the transit detections. Blended EBs that have primary and secondary stars with significantly different effective temperatures are ruled out, while those with similar effective temperatures could still possibly produce the transit signals based on the photometric data alone.

\subsection{Stellar Kinematics and Population}

Adopting the position, proper motion, parallax, and absolute radial velocity provided by \citet{GaiaDR2}, we compute the barycentric galactic velocity of \thisstar\, to be 
$U, V, W$ = -0.220$\pm$0.152, -45.803$\pm$0.181, -33.886$\pm$0.140 k\ms,
with $U$ measured towards the Galactic center, $V$ in the direction of  Galactic rotation, and $W$ towards the North Galactic Pole \citep[][]{ESA1997}. Adopting the Local Standard of Rest (LSR) from \citet{Schonrich2010}, this velocity translates to:
$U_{LSR}, V_{LSR}, W_{LSR}$ = 10.9, -33.6, -26.6 k\ms. 
Using the velocity ellipsoids and populations normalizations from \citet{Bensby2003},
we estimate the probabilities of kinematic membership of \thisstar\, to the thin disk,
thick disk, and halo to be 83.9\%, 16\%, and 0.1\%, respectively. Hence, TOI-824 is kinematically most consistent with being a thin disk star. The kinematic parameters
corroborate the chemical abundance information provided by \thisstar's color-magnitude diagram position (in the middle of the main sequence for field stars) and near-solar spectroscopic metallicity ([Fe/H] $\simeq$ -0.1), suggesting that TOI-824 is a typical thin disk star. 

\subsection{Stellar Multiplicity}

\begin{figure}
\includegraphics[width=.45 \textwidth]{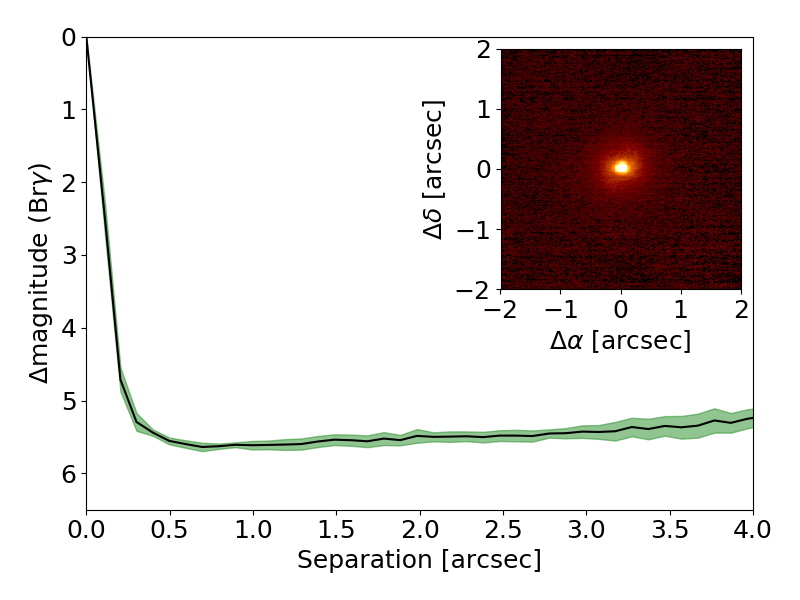}
\caption{VLT/NaCo contrast curve and image (inset) for TOI-824. Images were taken with a Br$\gamma$ filter and reach a contrast limit of 5 mag at 250mas. No visual companions are detected anywhere in the field of view.}
\label{fig:NACOplot}
\vspace{0.7cm}
\end{figure}

The combined NACO images show that no additional candidates were detected within the field of view, and that \target\ appears single to the limit of our resolution and contrast. The sensitivity of our observations was calculated as a function of radius by injecting fake companions, and scaling their brightness until they could be detected with 5$\sigma$ confidence. The contrast sensitivity is 5\,mag at 250\,mas, and 5.5\,mag in the wide field. The contrast sensitivity as a function of radius and a high resolution image of the star are shown in Figure \ref{fig:NACOplot}. The lack of companions strongly suggests that the transit signal originates from a planetary companion to TOI-824~b, rather than a background EB, and that the measured radius is not being diluted by a stellar companion \citep{Ciardi2015}.

We also searched for wide companions sharing similar proper motion and
parallax in the Gaia DR2 astrometric catalog \citep{GaiaDR2}. Any bound companions would likely be seen at separations smaller than the star's tidal (Jacobi) radius
($r_t$ $\simeq$ 1.35 pc ($M_{*}/M_{\odot})^{1/3}$), which for the
star's mass of 0.72 M$_{\odot}$, should correspond to about 1.21 pc
\citep{Mamajek2013,Jiang2010}, or 1$^{\circ}$.09 at the Gaia DR2
distance. 
A search of the Gaia DR2 catalog for stars with
parallaxes within 25\% of that of TOI-824 within 2$r_t$
(2$^{\circ}$.18) yielded 234 stars. 
Within projected separation of one tidal radius 
(1$^{\circ}$.09, 1.21 pc), {\it none} had a proper motion within 60 mas\,yr$^{-1}$ ($\Delta v_{tan}$ $\simeq$ 18 km\,s$^{-1}$) of that of TOI-824. 
Within two tidal radii, no Gaia DR2 stars had proper
motion within 20 mas\,yr$^{-1}$ ($\Delta v_{tan}$ $\simeq$ 6.1
km\,s$^{-1}$) of TOI-824. Additionally, among stars within two tidal radii,
no other Gaia DR2 candidates lacking parallaxes were found
with proper motions within $\pm$20 mas\,yr$^{-1}$ of that of TOI-824. 

The Gaia DR2 data for entries in the vicinity of TOI-824 are reasonably complete with both parallaxes and proper motions down to $G$ $\simeq$ 20.0 ($M_G$ = 16.0), and with increasing incompleteness down to $G$ $\simeq$ 21.4. Among nearby stars within 25 pc, absolute magnitude $M_G$ $\simeq$ 16 compares well to the M8.5V star 2MASS J11240487+3808054 (M$_G$ = 15.96, M$_{Ks}$ = 18.47; \citep{Cruz2003,Cutri2003,GaiaDR2}), whose $M_{Ks}$ value corresponds to mass 0.08 M$_{\odot}$ \citep{Mann2019}, just above the H-burning limit. So our search of the Gaia DR2 catalog for wide companions is likely complete to just above the H-burning limit or $\sim$0.08 $M_{\odot}$. 

Combining our high contrast imaging data, radial velocity data, and analysis of the Gaia DR2 astrometry for stars in TOI-824's vicinity, thus far the star appears to be a single star, although objects straddling the H-burning limit or brown dwarfs on wide orbits can not yet be ruled out. 

\subsection{Spectral Energy Distribution \label{sec:SED}}

\begin{figure}
\includegraphics[width=.45 \textwidth]{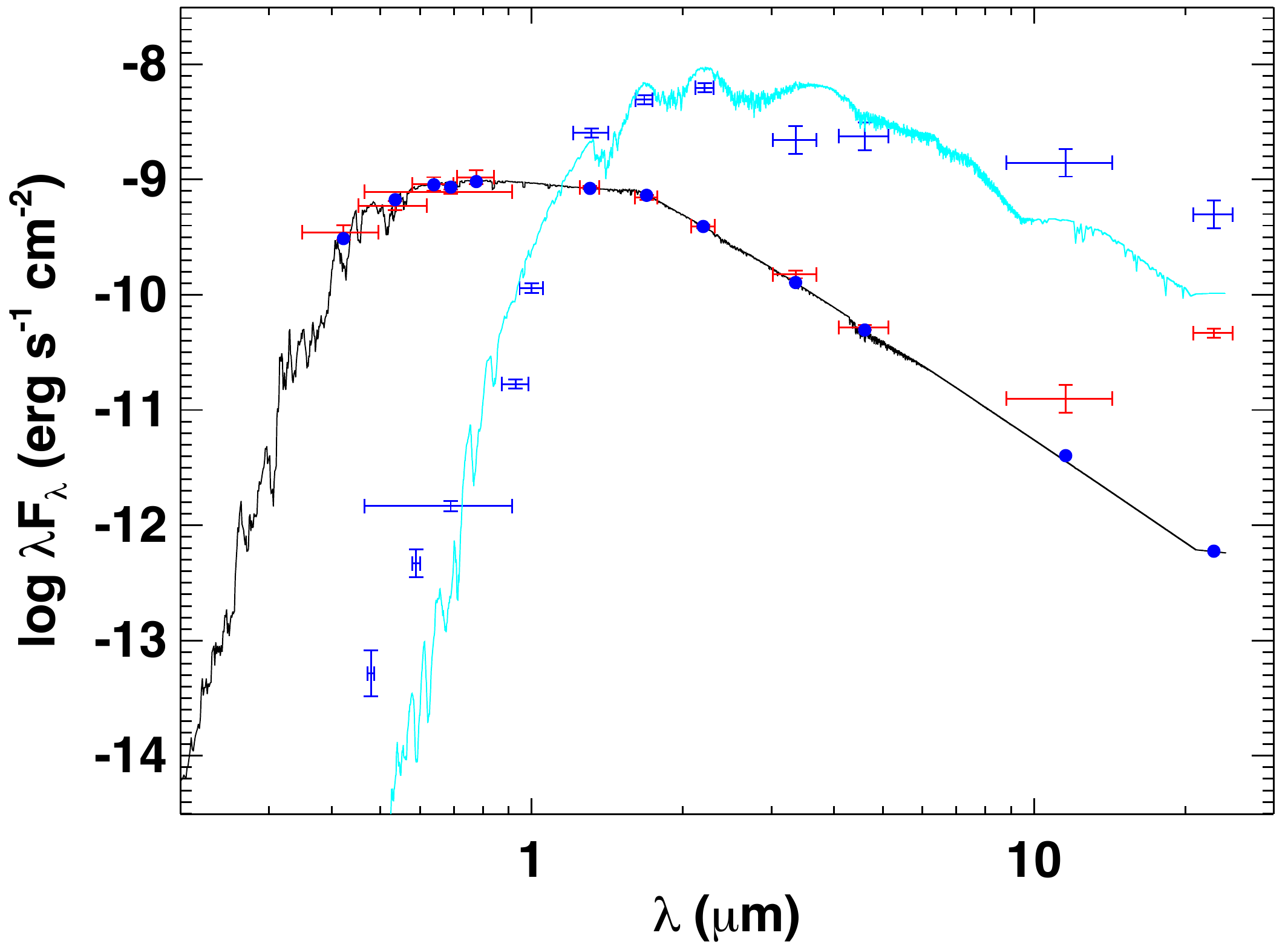}
\caption{Best fit spectral energy distribution (SED) for TOI-824. Red symbols represent the observed photometric measurements, where the horizontal bars represent the effective width of the passband. Blue symbols are the model fluxes from the best-fit NextGen atmosphere model, which is shown in black. A fit to the infrared-bright source separated by $\sim$25'' is represented by dark blue symbols and a cyan model. Note that the W3 and W4 bands (the two reddest) for TOI-824 were not used in the SED fitting, and the companion was ignored.}
\label{fig:SED_Fit}
\vspace{0.7cm}
\end{figure}

We performed an analysis of the broadband spectral energy distribution (SED) together with the {\it Gaia\/} DR2 parallax in order to determine an empirical measurement of the stellar radius, following the procedures described in \citet{Stassun:2016,Stassun:2017,Stassun:2018}. Together, the available photometry described in \S2\ and Table \ref{tab:star}, spans the full stellar SED over the wavelength range 0.4--22~$\mu$m (see Figure~\ref{fig:SED_Fit}). Noting the large excess in the {\it WISE\/3} and {\it WISE\/4} bands due to a nearby, infrared-bright star (IRAS 14448-5722, TIC 1133968082, Tmag=19.72), we chose to exclude them from the fit.  

We performed the fit using NextGen stellar atmosphere models, with priors set on the star's effective temperature ($T_{\rm{eff}}$), surface gravity ($\log g$), and metallicity ([Fe/H]) drawn from the TIC-8. We set the extinction ($A_V$) to zero due to the proximity of the star, which is consistent with the $A_{V} = 0.025 \pm 0.06$ value from \citet{Lallement2018}. The resulting fit is very good (Figure~\ref{fig:SED_Fit}) with a reduced $\chi^2$ of 2.3 and best-fit $T_{\rm eff} = 4450 \pm 100$~K. Integrating the model SED gives the bolometric flux at Earth of $F_{\rm bol} = 1.442 \pm 0.034 \times 10^{-9}$ erg~s~cm$^{-2}$. Taking the $F_{\rm bol}$ and $T_{\rm eff}$ together with the {\it Gaia\/} parallax, adjusted by $+0.082$~mas to account for the systematic offset reported by \citet{StassunTorres:2018}, gives the stellar radius as 
$R_{\star}$ = 0.719\,$\pm$\,0.033\,\rsun\, -- consistent with the updated stellar parameters listed in TIC-8.

In order to better estimate the potential flux contamination of the nearby, infrared-bright star $\sim$25" ($\sim$1 \TESS\ pixel) from TOI-824, we also performed an SED fit to that source. In this case we also fit for $A_V$, which we limited to the maximum line-of-sight extinction from the \citet{Schlegel:1998} dust maps. We used {\it Gaia\/} $G$, {\it 2MASS\/} $JHK_S$, {\it WISE\/} W1--W4, as well as $BV$ magnitudes from the SPM4.0 catalog \citep{Girard:2011} and $zy$ photometry from the VISTA catalog \citep{Cross:2012}. We obtained a best-fit $T_{\rm eff} = 2750 \pm 250$~K and $A_V = 5.8 \pm 1.9$. The {\it Gaia\/} DR2 parallax for this star is negative, so we instead used the Bayesian distance estimator from \citet{Bailer-Jones:2018} which together with integrated $F_{\rm bol}$ gives an estimated stellar radius of $R_{\star} = 1100 \pm 200$~R$_\odot$. The infrared-bright source is evidently a distant, highly extincted, red supergiant. We find that the brightness ratio between TOI-824 and this faint supergiant is $\sim$7000 in the TESS bandpass, and therefore conclude that it is unable to affect our measurement of the planet's radius at a detectable level given the final radius error bars of $\pm 0.1$ \rearth.

\subsection{Stellar Parameters \label{sec:StellarParams}}

We matched the CHIRON spectra against a library of $\sim 10,000$ observed spectra classified by the Stellar Parameter Classification (SPC) pipeline \citep{Buchhave:2012}, interpolated via a gradient boosting regressor. From this analysis, we find the effective temperature, metallicity, surface gravity, and rotational velocity of \target\ to be: $T_\mathrm{eff} = 4665 \pm 100$\,K, $\log g = 4.68 \pm 0.10$ dex, $\mathrm{[m/H]} = -0.28\pm0.10$ dex, and $v\sin i = 4.5 \pm 0.5\,\mathrm{km\,s}^{-1}$, all of which suggested that \target\ was suitable for precision radial velocity follow up efforts.

To determine more precise constraints of the stellar parameters for \thisstar, which have a large influence on the derived planetary parameters, we took one of the spectra from HARPS (which is higher resolution than the spectra provided by CHIRON) and analyzed it using the SPC pipeline described above. From the HARPS spectrum, SPC reports that the star has an effective temperature T$_{\rm eff} = 4569\, \pm\, 50$ K, a surface gravity of $\rm{log}(g) = 4.56 \pm\ 0.10$, and a metallicity of [m/H]$ = -0.12 \pm\ 0.08$. We performed a secondary check by running the same HARPS spectra through the SpecMatch-emp software package \citep{Yee2017} which reported similar results that produced minimal changes when used to derive planetary parameters for \thisstar~b. Given this general agreement, we adopt the HARPS + SPC stellar parameters as the priors for our final EXOFASTv2 analysis of the combined data sets.

\subsection{Individual Stellar Abundances \label{sec:species}}

To determine the abundances of specific elements in TOI-824 we combined the individual HARPS spectra into a single, high SNR spectrum and applied the \texttt{SPECIES} code \citep{Soto2018}. SPECIES computes the atmospheric parameters (\teff, \logg, \feh, v$_{\rm{t}}$) parameters by measuring the equivalent widths (EW)\footnote{The equivalent widths were computed using the \texttt{EWComputation} module available at https://github.com/msotov/EWComputation.} for a set of iron lines. These, together with an ATLAS9 model atmosphere \citep{ATLAS9}, are used to solve the radiative transfer equation in the atmosphere of the star using MOOG \citep{moog}. Abundances for individual ions are estimated by computing the EWs for a set of lines and using the derived parameters from before to create an appropriate atmospheric model to input to MOOG. Physical parameters, including stellar mass and radius, were obtained by interpolating through a grid of MIST models \citep{Dotter2016}, using the isochrones python module \citep{Morton2015}. The atmospheric parameters, along with the magnitude of the star at different filters and its parallax (Table~\ref{tab:star}), were used as priors for the interpolation. Finally, the macroturbulence velocity was obtained from the effective temperature, and the projected rotational velocity by broadening the profiles of a set of absorption lines. The lines used in the fitting procedure, along with the absorption lines used in the abundance determination, are listed in \citet{Soto2018}. The results from SPECIES are shown in Table~\ref{tab:species}.

\begin{table}[htbp]
\centering
\caption{SPECIES results for TOI-824}
\begin{tabular}{lcc}
\hline\hline
Parameter	& Value & Uncertainty \\
\hline \vspace{2pt}
[Fe/H] [dex]	         & -0.15 & 0.02\\
$T_\mathrm{eff}$ [K] &	4616 & 51\\	
log\,$g$ [[cm s$^{-2}$]]	     &  4.613 & 0.12\\ 
$v_t$ [k\ms] &	0.188 & 0.10\\
$v$sin$i$ [k\ms]    &  2.165 & 0.21\\
$v_\mathrm{mac}$ [k\ms] & 1.542 & 0.02\\
\# Fe\,I lines   & 131 & \\
\# Fe\,II lines  &	 8 & \\
\hline\hline \vspace{2pt}
Element &   Value   & \# Lines \\
\hline
{[Na/H]} &	0.08$\pm$0.20   & 1\\ 
{[Mg/H]} &	-0.23$\pm$0.12	& 3\\ 
{[Al/H]} &	-0.18$\pm$0.12	& 3\\ 
{[Si/H]} &	-0.18$\pm$0.12	& 3\\ 
{[Ca/H]} &	-0.48$\pm$0.08	& 7\\ 
{[Ti\,I/H]}  & 0.00$\pm$0.08 & 6\\ 
{[Ti\,II/H]} & 0.27$\pm$0.14 & 2\\ 
{[Cr/H]} &	-0.12$\pm$0.06	& 12\\ 
{[Mn/H]} &	0.03$\pm$0.09	& 5\\ 
{[Ni/H]} &	-0.04$\pm$0.09	& 5\\ 
{[Cu/H]} &	0.39$\pm$0.12	& 3\\ 
{[FeI/H]} & -0.07$\pm$0.06 & 13\\
{[FeII/H]} & -0.02$\pm$0.10 & 4\\
\hline\hline \vspace{2pt}
Parameter & Value & 54\% Confidence Level\\
\hline
Mass [\msun]          &	0.69  & $^{0.009}_{0.007}$ \\		
Age	[Gyr] &	10.9 & $^{1.8}_{3.1}$ \\
log g$_{iso}$ [c\ms] &	4.612 & $^{0.011}_{0.007}$ \\		
Radius [\rsun] & 0.68 &	0.005 \\		
log(L/L$_{\odot}$) & -0.72 & $^{0.009}_{0.012}$ \\
\hline\hline
\end{tabular}
\label{tab:species}
\end{table}

\subsubsection{Upper Age Constraint from [$\alpha$/Fe]}

Based on the measured abundances of the $\alpha$ elements 
Mg, Si, Ca, Ti (using only Ti I) and the
Fe abundance, we calculate the $\alpha$ enrichment [$\alpha$/Fe].
We follow \citet{Bovy2016} and calculate a mean $\alpha$ abundance using Mg, Si, Ca, and Ti\,I, however we omit O and S, which were not measured.
Weighting by the number of lines used for the abundance
of each species, we estimate [$\alpha$/H] = -0.19.
Given the star's iron abundance [Fe/H] = -0.15\,$\pm$\,0.018, 
this translates to [$\alpha$/Fe] $\simeq$ -0.04. 
From comparison of the [Fe/H] and [$\alpha$/Fe] estimates to those of local age-dated FGK stars in the survey of \citet{Haywood2013}, it appears that stars with TOI-824's metallicity and solar $\alpha$ abundances are all thin disk stars with isochronal ages of $\lesssim$8 Gyr.
While there are older stars with ages $\sim$8-10 Gyr classified as thin disk, they tend to be more metal poor and more $\alpha$-rich \citep{Haywood2013}.
The kinematic data also supports classification of TOI-824 as a likely thin disk star, and its membership to the thin disk provides an independent age constraint (95\%CL upper limit) of $<$8 Gyr. 

\subsubsection{Lower Age Constraint from the Li 6707\AA\ Line}

Analysis of the Lithium 6707.8\AA\ line region in the HARPS spectra shows no signs of the absorption feature. We are able to place a strong 10 m\AA\ upper limit on the line's equivalent width, indicating that TOI-824 is a Li-poor K dwarf. For T$_{\rm{eff}} \simeq 4600K$, a Li 6707 EW \textless\ 10m\AA\ is inconsistent with (i.e. older than) M7/NGC6475 \citep[age=220Myr,][]{Sestito2003} and M34 \citep[age=250 Myr,][]{Jones1997}. That Li EW is consistent with the mixture of detections and non-detections of Li 6707 in the Hyades \citep[age=700Myr,][]{Barrado1996} and the Praesepe Cluster \citep[age 590-790 Myr,][]{Cummings2017}. These comparisons show that TOI-824 is almost certainly older than 250 Myr, and likely older than 500Myr.

\subsection{System Parameters from EXOFASTv2}
\label{sec:exofastv2}

To fully characterize the \thisstar\ system, we used the EXOFASTv2 software package \citep{Eastman2017, Eastman2019} to perform a simultaneous fit to the \TESS\ photometry, the ground based SG1 photometry, and the radial velocities from PFS and HARPS. The detrending of the ground-based photometry, to correct for observational effects such as changes in the star's airmass throughout the transit, is handled within EXOFASTv2 using parameters provided by the AIJ software.

\subsubsection{EXOFASTv2 Priors and Starting Values}

We enforced Gaussian priors on the star's effective temperature (T$_{\rm eff} = 4569\pm114$ K) and metallicity ([Fe/H] = $-0.12\pm0.08$) using the SPC analysis results of the HARPS spectrum described in Section \ref{sec:StellarParams}, and on the stellar radius (R$_{\star} = 0.719\pm0.0333$ $R_{\odot}$) using the results of the SED fit described in Section \ref{sec:SED}. We also placed a Gaussian prior on the star's parallax from Gaia's DR2 results ($\pi = 15.696178\pm0.04934$ mas) after applying the correction from \citet{StassunTorres:2018}. All starting values were further refined using the results of earlier, shorter, EXOFASTv2 fits. 
 
The orbit of planet b was defined to be circular in our analysis, as initial EXOFASTv2 fits to the data found eccentricity values consistent with zero and previous studies of small, short-period planets have generally found low eccentricities \citep{HaddenLithwick2017,VanEylenAlbrecht2015}. We also allowed for a linear slope to be applied to the RV data during the fitting process. The EXOFASTv2 RV model fits for velocity offsets between the PFS and HARPS data sets as well as different instrumental jitter values, terms that are added in quadrature to the estimated measurement uncertainties from PFS and HARPS to account for systematic effects. To constrain the star's age, we use EXOFASTv2's implementation of the MESA Isochrones and Stellar Tracks (MIST) stellar evolution models \citep{Paxton2013, Paxton2015, Dotter2016, Choi2016}.

We also fit a dilution term to the \TESS\ photometry to check whether additional correction is needed to address blending from nearby stars. The dilution factor of the \TESS\ photometry is determined by comparing the \TESS\ transit depth to the transit depth measured in the ground based light curves. The ground based photometric data has higher spatial resolution and therefore we expect these transits to experience less flux contamination. We performed two instances of the EXOFASTv2 fit. In the first, which is the version that we use for the final planet parameters presented in Table \ref{tab:EXOFASTv2}, the \TESS\ dilution parameter is unconstrained. In the second, we enforce a Gaussian prior of 0.0$\pm$0.03 on the \TESS\ dilution. If the SPOC pipeline that produced the \TESS\ light curves corrected the blending effects properly then the best fit to the dilution parameter should be close to zero, regardless of fitting priors.

\subsubsection{EXOFASTv2 Results}

\begin{figure}
\includegraphics[width= .98 \linewidth]{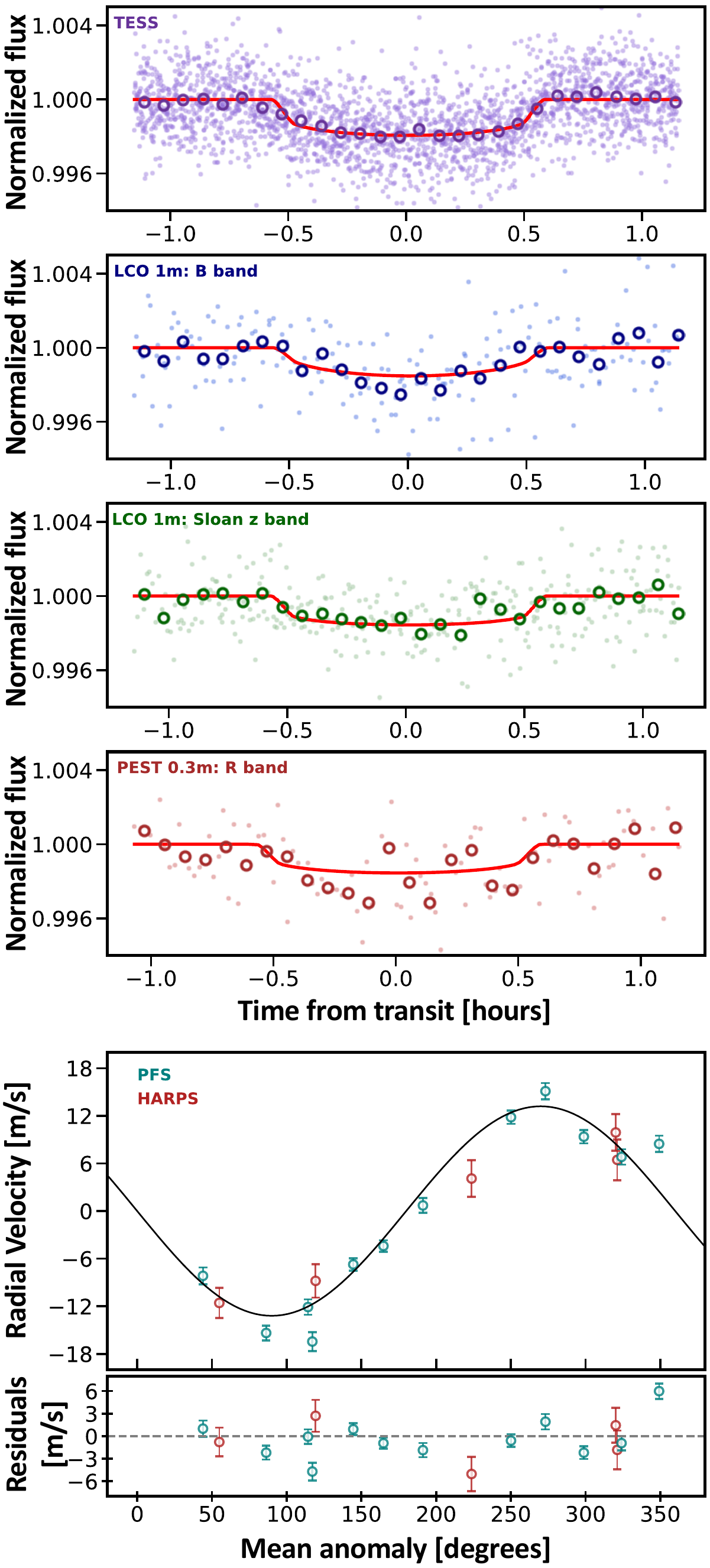}
\caption{Results of the EXOFASTv2 joint fit to the detrended photometry from TESS, LCO, and PEST (top panels) and the radial velocity measurements from PFS and HARPS (bottom panel). Light points in the photometry panels depict the unbinned observations, while darker circles show the data in 30-minute bins. Solid lines show the best fit models to both the photometry and RV data sets. The planet's transit has a depth of 1490 ppm and a total duration of 1.15 hours, while the RV curve has a semi-amplitude of 13.2 \ms. 
\label{fig:CombinedFit}}
\vspace{0.7cm}
\end{figure}

\begin{figure}
\includegraphics[width= .98 \linewidth]{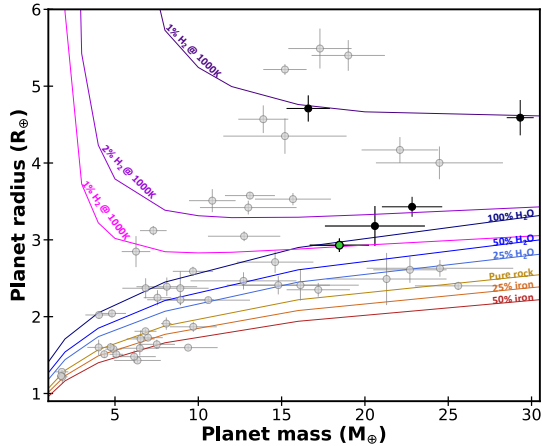}
\caption{Mass-radius diagram for confirmed planets with masses and radii measured to better than 25\% (gray points) retrieved from the Exoplanet Archive (https://exoplanetarchive.ipac.caltech.edu). Black points show the other hot Neptune planets recently discovered by \TESS\ and NGTS. TOI-824b is denoted with a green circle. Composition curves from \citet{Zeng2016} and \citet{Zeng2019} are shown as solid, colored lines.
\label{fig:MRDiagram}}
\vspace{0.7cm}
\end{figure}

The median EXOFASTv2 parameters for the \thisstar\ system are shown in Table \ref{tab:EXOFASTv2} and the best fits to the TESS photometry and PFS and HARPS radial velocity data are shown in Figure \ref{fig:CombinedFit}. The mass of \thisplanetb\ is measured to be \mplanetunc\ which, when combined with the measured planet radius of \rplanetunc\, results in a bulk density of \rhoplanetunc, making the planet more than twice as dense as Neptune (Figure \ref{fig:MRDiagram}). This radius measurement is roughly 15\%\ smaller than the R = 3.4 \rearth\ estimate based upon the \TESS\ data alone (Section \ref{sec:transitdetection}) and we address this difference below in Section \ref{sec:dilutioneffects}. We find that our assumption of a circular orbit is further supported by the fact that the tidal circularization timescale ($\tau_{\rm circ} \sim\,0.57$ Gyr) is short compared to the star's age ($7.5^{+1.8}_{-2.9}$ Gyr). 

\begin{figure}
\includegraphics[width=.45 \textwidth]{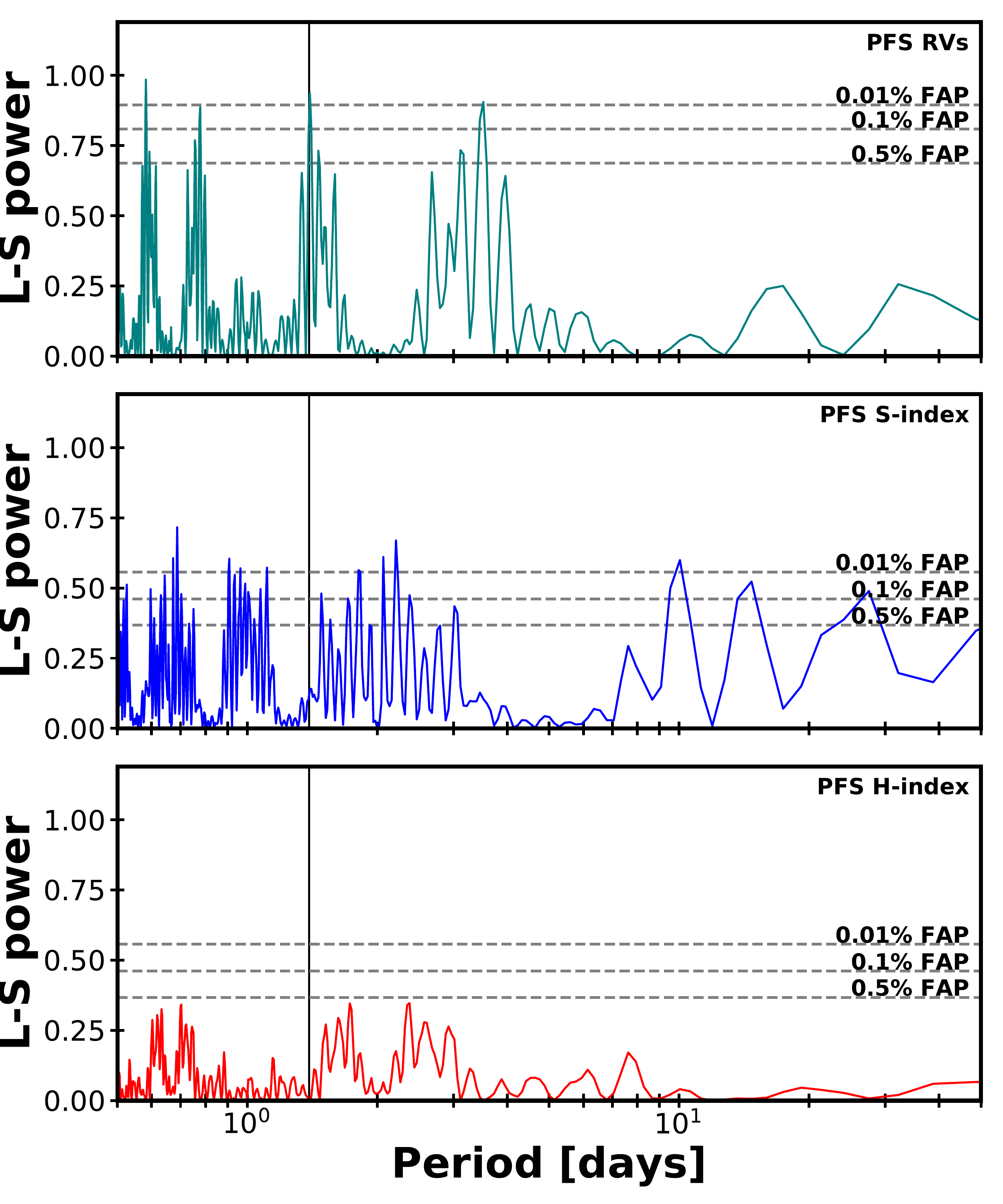}
\caption{Lomb-Scargle periodograms of the radial velocities (top), S-index values (middle), and H-index values (bottom) extracted from the PFS spectra. The period of TOI-824~b is marked by a vertical line in each panel, and the 0.5\%, 0.1\%, and 0.01\% False Alarm Probabilities are marked with dashed, horizontal lines. The signal of the planet is clearly visible in the top most panel, but neither of the activity indicator periodograms display significant peaks in the same region of period space.
\label{fig:Periodograms}}
\vspace{0.7cm}
\end{figure}

For the sake of completeness we check the S-index and H-index activity indicators extracted from the PFS data set for periodicities that could cause the 1.39-day signal. These activity indicators serve as as proxies for chromospheric activity in the visible stellar hemisphere at the moments when the spectra were obtained. The S-index is calculated by measuring the emission reversal at the cores of the Fraunhofer H and K lines of Ca II located at at 3968 {\AA} and 3934 {\AA}, respectively \citep{Duncan1991}, while the H-index quantifies the amount of flux within the H$\alpha$ Balmer line core compared to the local continuum. Details on the prescription used for measuring these indicators in the PFS data set can be found in \citet{Butler2017}. We analyze the resulting S- and H-index values by computing Lomb-Scargle periodograms for each of the activity indicators as well as for the PFS radial velocity values and then looking for any well defined peaks with False Alarm Probabilities $<$0.1\%\, in the vicinity of the planet's period (Figure \ref{fig:Periodograms}). 

The RV signal of TOI-824~b is apparent in the Lomb-Scargle periodogram of the PFS radial velocities (Figure \ref{fig:Periodograms}). Neither the S- nor H-index periodogram displays any significant peaks at periods close to the planetary signal, however, which means stellar activity is unlikely to skew or otherwise influence our measurement of the planet's mass. Indeed, based upon these periodograms, TOI-824 appears to be a relatively quiet star. 

Given the match in both period and phase between the \TESS\ signal and the planet signal seen in the combined RV data set, along with the lack of any significant periodicity in the spectral activity indicators, we consider this to be a decisive confirmation of the planetary nature of \thisplanetb.


\section{Correction to \TESS\ Background Flux Estimates \label{sec:dilutioneffects}}
The final EXOFASTv2 fit finds a significant, negative dilution factor (A$_{D}$ = -0.26) for the \TESS\ photometry. This dilution value is determined by comparing the $\sim$1888ppm depth of the \TESS\ transits to the 1490ppm depth of the ground-based transits and a negative A$_{D}$ value indicates that the \TESS\ light curve has a higher out-of-transit flux than is reported by the SPOC pipeline. The \TESS\ SPOC data therefore exhibits larger fractional flux drops during transit events, leading to a larger measured planet radius. In the case of TOI-824~b, including the \TESS\ dilution value in the EXOFASTv2 fit results in a 13\% decrease in the measured planet radius when compared to earlier fits that used only the \TESS\ data to measure R$_{p}$ and not the ground-based follow up data.

\begin{figure}
\includegraphics[width=.48\textwidth]{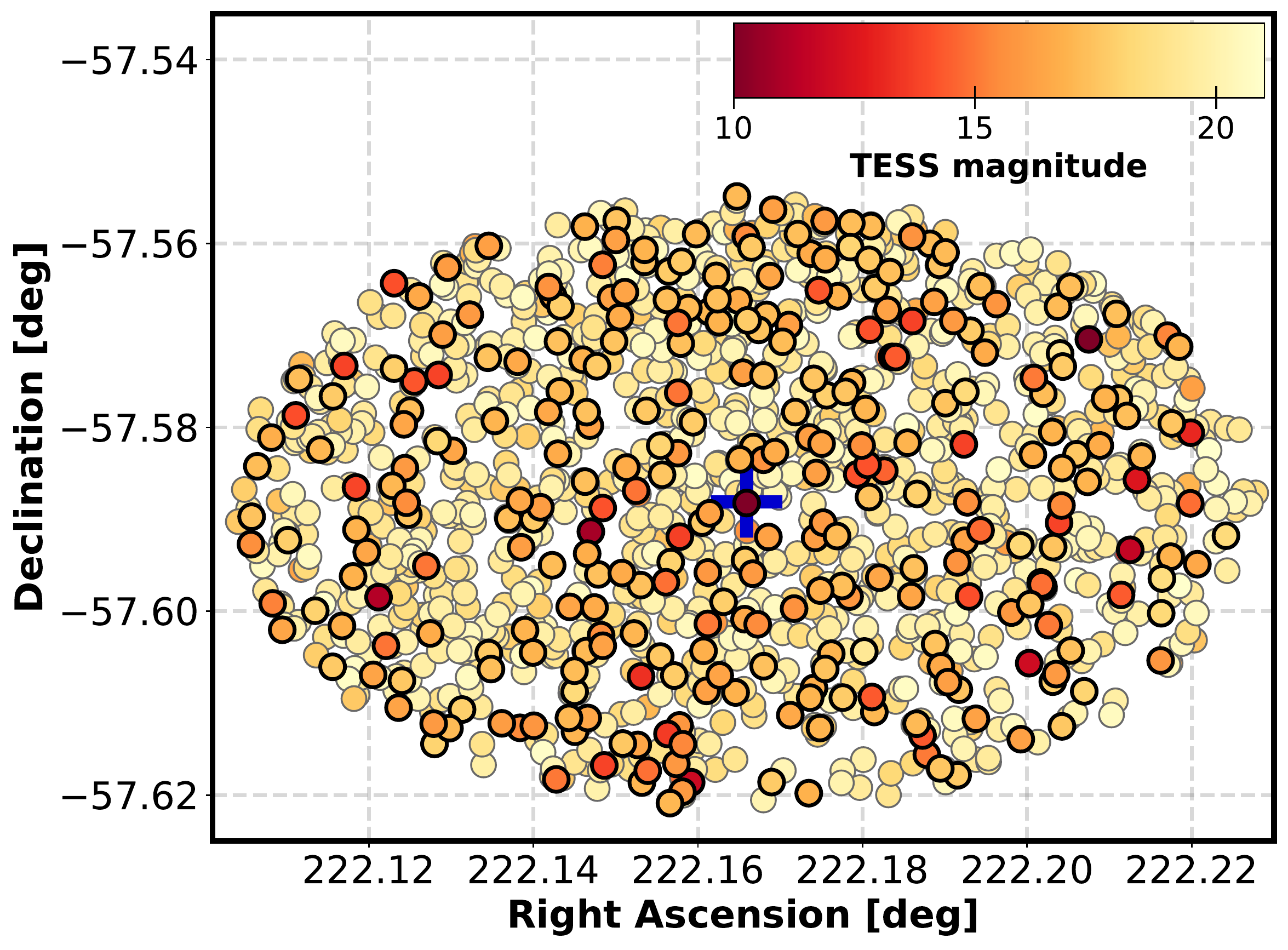}
\caption{Stars present in TICv7 (black outlines) and TICv8 (grey outlines) within 2 arc minutes of TOI-824 which is identified with a blue cross. All TICv7 stars are also present in TICv8, along with an additional 1049 stars identified by Gaia DR2. The vast majority of these additional stars are more than 8 TESS magnitudes fainter than TOI-824 and so we do not anticipate them causing problems with the crowding correction. 
\label{fig:TIC_Compare}}
\vspace{0.7cm}
\end{figure}

We first investigated whether this offset could be caused by an inaccurate correction in the \TESS\ data for the effects of nearby stars, as TOI-824 is in a crowded part of the sky and numerous additional sources fall within the SPOC aperture. We find, however, that this is not the case. The TESS SPOC pipeline includes a crowding correction to address the effects of flux from nearby sources. This correction is based on a simulated star scene for each CCD that is created using the detectors' Pixel Response Functions (PRFs, measured during commissioning) and stars in the TIC catalog that are imaged by the CCD. The simulated star scene is used to estimate the fraction of flux within a given TOI’s photometric aperture that is due to the target star compared to the total flux contributed by all stars whose images fall within the photometric aperture. That ratio is then used to correct the resulting SPOC light curve. Like all stars observed in Sectors 1-13, TOI-824's crowding correction is based upon TICv7 which uses the Gaia DR1 and 2MASS catalogs among others. The more recent TICv8, however, is based upon the Gaia DR2 catalog and identifies an additional 1049 stars within that same radius (Figure \ref{fig:TIC_Compare}). While this is a large increase in nearby stellar neighbors, 99.6\% of the newly identified TICv8 stars are fainter than T$_{\rm{mag}}$ = 15 (five TESS magnitudes fainter than TOI-824) and 91\% are fainter than T$_{\rm{mag}}$ = 18. But accounting for these additional faint TICv8 stars, present in the \TESS\ images but not included in the crowding correction, would increase the deblended SPOC depth and resulting planet radius rather than decrease it, exacerbating the issue.

\begin{figure}
\includegraphics[width=.48\textwidth]{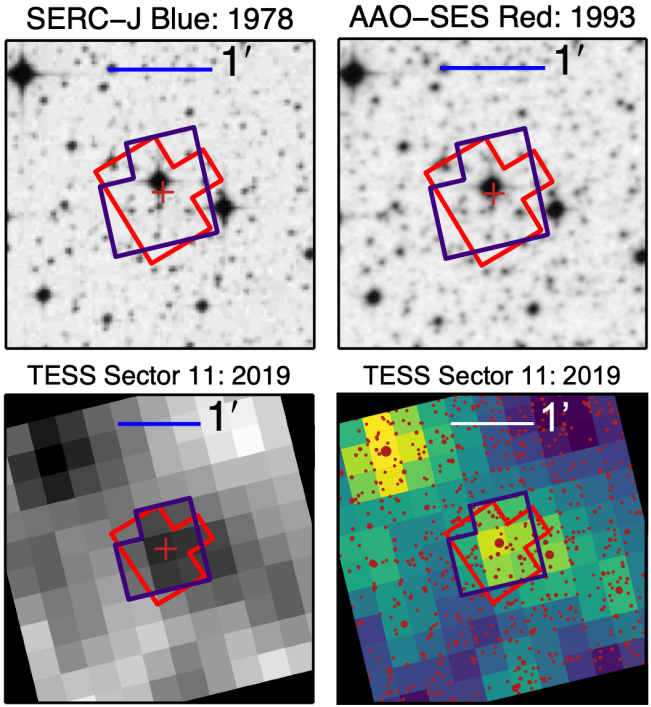}
\caption{Images of the crowded star field surrounding TOI-824. The red cross in each image is the present day location of TOI-824 and the purple and red lines mark the boundary of the TESS photometric apertures for Sectors 11 and 12, respectively. North is up and East is to the left in all the images. \emph{Top left}: SERC-J survey, wavelength coverage from the UV to 540 nm. \emph{Top Right}: ESO/SERC Southern Sky Atlas, wavelength coverage of 590 - 690 nm. \emph{Bottom Left}: Summed TESS Sector 11 image. \emph{Bottom Right}: Single TESS Sector 11 frame with Gaia DR2 sources depicted as maroon circles.
\label{fig:PostageStamps}}
\vspace{0.7cm}
\end{figure}

The discrepancy comes instead from an overestimation of the TESS background flux caused by the plethora of nearby stars, many of which fall into TOI-824's postage stamp aperture (Figure \ref{fig:PostageStamps}). When examining the pixels that make up the Sector 11 postage stamp for TOI-824 we find that the dimmest background-corrected pixels have a median value of roughly -180 e$^{-}$/second. This suggests that the background level measured in the postage stamp's ``empty" pixels -- which, given the high stellar density of the region, are not truly empty -- was too high and its subtraction from the TOI-824 aperture produced negative flux measurements in some pixels. If we adjust the mean flux of the postage stamp upwards by 8 x 180 e$^{-}$/sec to account for the 8 pixels in the optimal aperture then the mean flux becomes 13640 e$^{-}$/sec. This reduces the SPOC transit depth from 1888 ppm to 1663 ppm for Sector 11, a 12\% decrease which produces a 6\% reduction in the planetary radius.

The ground-based photometry, however, suggests an even smaller transit depth of 1490 ppm which is 21\% below the original SPOC transit depth and another 10\% below the corrected value outlined above. Given that no pixel within the TOI-824 postage stamp is completely devoid of stars, making an accurate estimate of the background flux nigh on impossible, we find it to be very plausible that an additional factor of 10\% overestimation could be folded into the SPOC light curves. To investigate this possibility, we examined a 51 x 51 pixel FFI cut out image from Sector 11 centered on TOI-824. After examining the star's FFI time-series, we selected an image that is minimally contaminated with scattered light, which turned out to be the last FFI image taken in Sector 11. We identified the darkest 40 pixels in the FFI image and calculated an average background of 325 e-/sec, which resulted in an adjusted SPOC Sector 11 depth of 1517 ppm. This adjusted depth is now within 2\% of the ground-based depth, verifying that it is indeed the background flux estimation causing the \TESS\ pipeline to report an inflated planet radius.

The radius offset between the \TESS\ and SG1 data for TOI-824 b highlights the importance of inspecting \TESS\ data products at the pixel level before using them to determine planet characteristics. And it reinforces even more so the critical role that ground-based follow up transit observations and fitting approaches that allow for offsets between independent light curves play in correctly measuring planet radii. While this is most relevant when considering stars in crowded regions, as in the case of TOI-824, the general principle holds for almost any exoplanet science derived from \TESS\ data. Similar effects from incorrect background corrections have been seen in both \citet{Kostov2020} and \citep{Feinstein2020}. \citet{Kostov2020} corrects the offset by using the \TESS\ FFIs to infer an appropriate background flux which then gets added back into the 2-minute cadence data, similar to our approach using the ground based SG1 photometry. In comparison, \citep{Feinstein2020} determines their background estimate by analyzing a given star's entire postcard region (148x104 pixels) and then subtracting the resulting background flux before extracting the target pixel files which are in turn used to produce light curves.

We expect the background subtraction bias seen here to predominantly affect dim stars or stars in highly crowded regions. Indeed, when investigating all 2-minute targets in Sector 14, which included the plane of the galaxy, we find that if all stars hosted transiting planets then the change in planet radius due to background bias would be less than 1\% for 70\% of cases. In Sector 22, which does not include the plane of the galaxy and therefore has less stellar crowding, if we again assume that all targets host transiting planets then the percentage of planets affected at the \textless 1\% level rises to 87\%. Thus for for most objects of interest the change in planet radius due to background bias will likely be much smaller than other sources of error in planet radius.

TOI-824, which sits in a very crowded region of the sky, is one of the strongest background bias cases detected to date with a 10.5\% planet radius reduction. In response to this issue, the SPOC has updated the background estimation algorithm to prevent background-subtracted pixel time series from being significantly negative, and will begin applying it starting with Sector 27. As a general guide line we recommend that when working with \TESS\ photometry scientists should, whenever possible, incorporate some additional measure of the transit depth into their analysis instead of relying solely on the two-minute photometry. These additional analysis measures will become especially important when \TESS\ begins its observations of the more crowded ecliptic equator in Cycle 4.

\section{Discussion}
\label{sec:discussion}

\subsection{Interior Characterization TOI-824 b}
\label{sec:interior}
We model the interior of TOI-824 assuming a pure iron core, a silicate mantle, a pure water layer, and a H-He atmosphere. We follow the structure model of \citet{Dorn2017}, with the EOS of the iron core taken from \citet{Hakim2018}, the EOS of the silicate-mantle calculated using PERPLE\_X by \citet{Connolly2009} given thermodynamic data of \citet{Stixrude2011} and  \citet{Saumon1995} for the H-He envelope assuming protosolar composition. For the water we use the quotidian equation of state (QEOS) presented in \citet{Vazan2013} for low pressures and the tabulated EOS from \citet{Seager2007} for pressures above 44.3 GPa. We then use a generalized Bayesian inference analysis using a Nested Sampling scheme \citep[e.g.][]{Buchner2014} to quantify the degeneracy between interior parameters and produce posterior probability distributions.  We use the stellar Fe/Si and Mg/Si ratios as a proxy for the planet, and assume an envelope luminosity of L=$10^{22.52}$ erg/s (equal to Neptune's luminosity). \\

\begin{table}[htbp]
\centering
\caption{Inferred Interior Structure Properties of TOI-824b.}
\label{tab:InteriorComp}
\begin{tabular}{ll}
\hline\hline
$M_{core}/M_{total}$ & $0.27^{+0.23}_{-0.11}$\\
$M_{mantle}/M_{total}$ & $0.38^{+0.25}_{-0.18}$\\
$M_{water}/M_{total}$ & $0.31^{+0.24}_{-0.18}$\\
$M_{atm}/M_{total}$ & $0.028^{+0.008}_{-0.007}$\\
\hline\hline
\end{tabular}
\end{table}

Table \ref{tab:InteriorComp} lists the inferred mass fractions of the core, mantle, water-layer, and H-He atmosphere from our structure models. We find a median H-He mass fraction of 2.8\%, which is a lower-bound since enriched H-He atmospheres  are more compressed, and can therefore increase the planetary H-He mass fraction. Indeed, formation models of mini-Neptunes suggest that it is very unlikely to form such planets without envelope enrichment \citep{VenturiniHelled2017}. The core, mantle, and water layer have relative mass fractions between 27\%, 38\% and 31\% with large sigma. This regime of the M-R relation is very  degenerate, and therefore it is not possible to accurately determine the  mass ratios of the core, mantle, and water layer. 

\subsection{Structure and atmospheric evolution of TOI-824 b}
\begin{figure}
\includegraphics[width=0.98\linewidth]{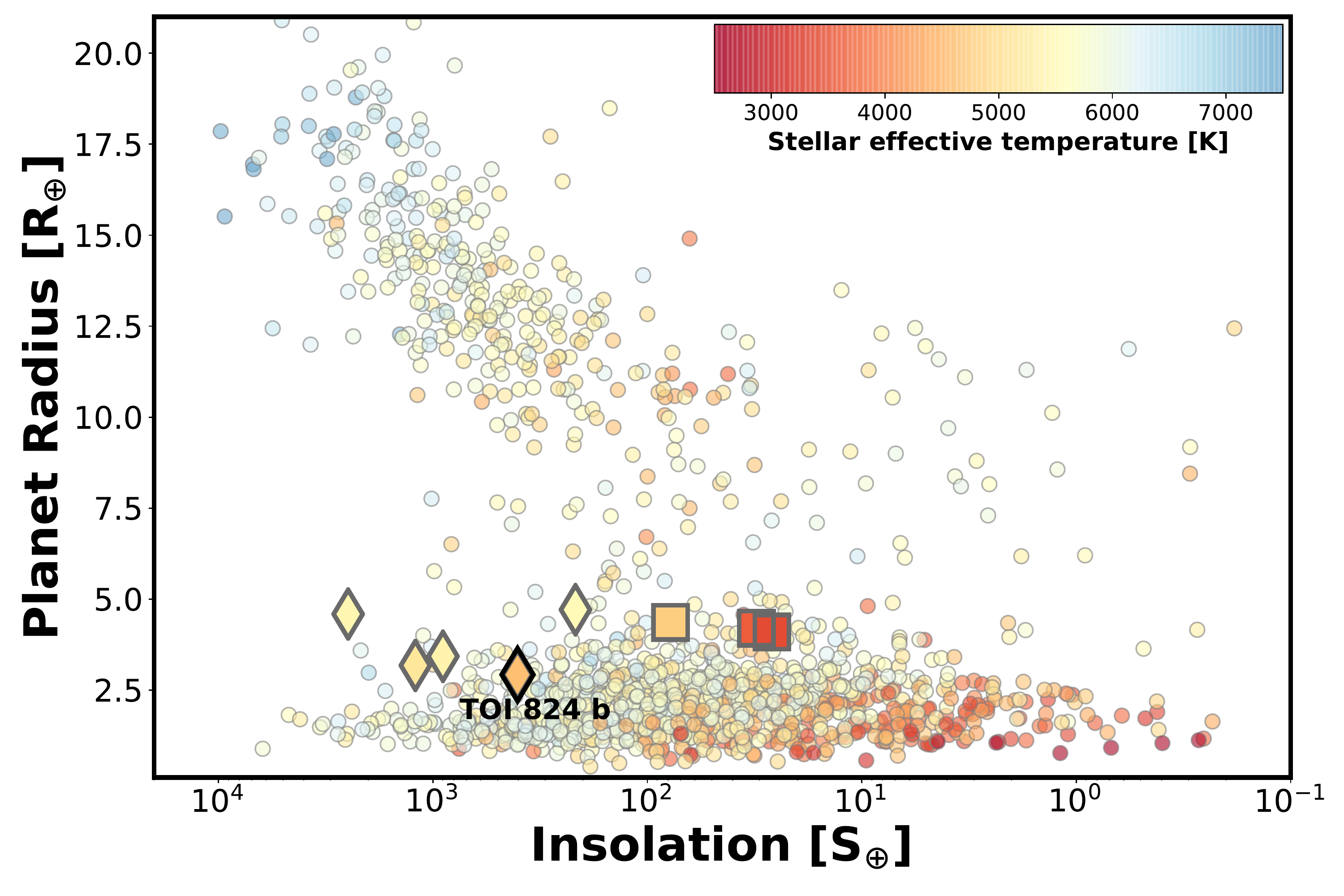}
\caption{\object{TOI-824 b} (black diamond) sits on the lower edge of the hot Neptune Desert. Other hot Neptunes discovered within the past year (\object{TOI 132 b}, \object{LTT 9779 b}, \object{NGTS-4b}, and \object{HD 219666 b}) are shown as diamonds, while comparable planets with well-studied atmospheres (\object{HAT-P-11b}, \object{GJ 3470 b}, and \object{GJ 436 b}) are shown as squares. \object{TOI-824 b} inhabits a notably different region of this parameter space than even the most irradiated of the planets with well studied atmospheres (\object{HAT-P-11b}) and offers an opportunity to investigate how increased irradiation affects a variety of atmospheric characteristics.
\label{fig:DesertFig}}
\vspace{0.7cm}
\end{figure}

One of the most intriguing results of NASA's \Kepler\ mission is clear evidence that the overall distribution of small, short-period, planets has been sculpted by processes that erode atmospheres \citep[e.g.,][]{Lopez2012,Owen2013,Chen2016,Owen2017,Jin2018}. This evidence includes both the clear gap in the planet radius distribution uncovered by \citet{Fulton2017T} and better documented in \citet{Fulton2017} and \citet{Fulton2018}, as well as the clear dearth of non-rocky 2-4 \Rearth\, planets in the most strongly irradiated orbits \citep[e.g.,][]{Sanchis-Ojeda2014,Lundkvist2016,McDonald2019}, which is frequently referred to as the hot Neptune Desert. This desert is normally shown by examining the distribution of planetary radii and insolations, as in Figure \ref{fig:DesertFig}, however this is closely related to similar concepts like the ``Cosmic Shoreline" described in \citet{Zahnle2017} which compares planetary insolation and escape velocity, as well as the mass-loss thresholds found by comparing planetary binding energies to the high ionizing X-ray and EUV irradiation they receive \citep[e.g.,][]{Lecavelier2007,Lopez2013,Owen2013,Lopez2014}. Indeed the hot Neptune Desert and the radius gap closely match prior predictions from models of extreme atmospheric escape due to XUV driven photo-evaporative escape \citep[e.g.,][]{Owen2012,Lopez2013,Owen2013,Jin2014,Lopez2017}, although other extreme escape mechanisms have subsequently been proposed to explain these features \citep[e.g.,][]{Schlichting2015, Ginzburg2018}.

\object{TOI-824 b} is particularly interesting in the context of the hot Neptune Desert since, along with a handful of other recent discoveries, it appears to lie at the lower edge of the desert (see Figure \ref{fig:DesertFig}). Its mass and radius, however, indicate that \object{TOI-824 b} must possess a significant primary atmosphere. Assuming a rock and iron core, thermal evolution models from \citet{Lopez2014} suggest a H+He envelope fraction of 2.4$^{+1.1}_{-1.7}$\%, consistent with the findings in Section \ref{sec:interior}. This is well within the typical range of the warmest Neptune planets discovered by \Kepler, although of course those are typically much less irradiated than \thisplanetb. This poses an interesting question: how could this planet have possibly retained a significant gaseous envelope despite receiving extreme radiation? 

\begin{figure}
\includegraphics[width=1.0\linewidth]{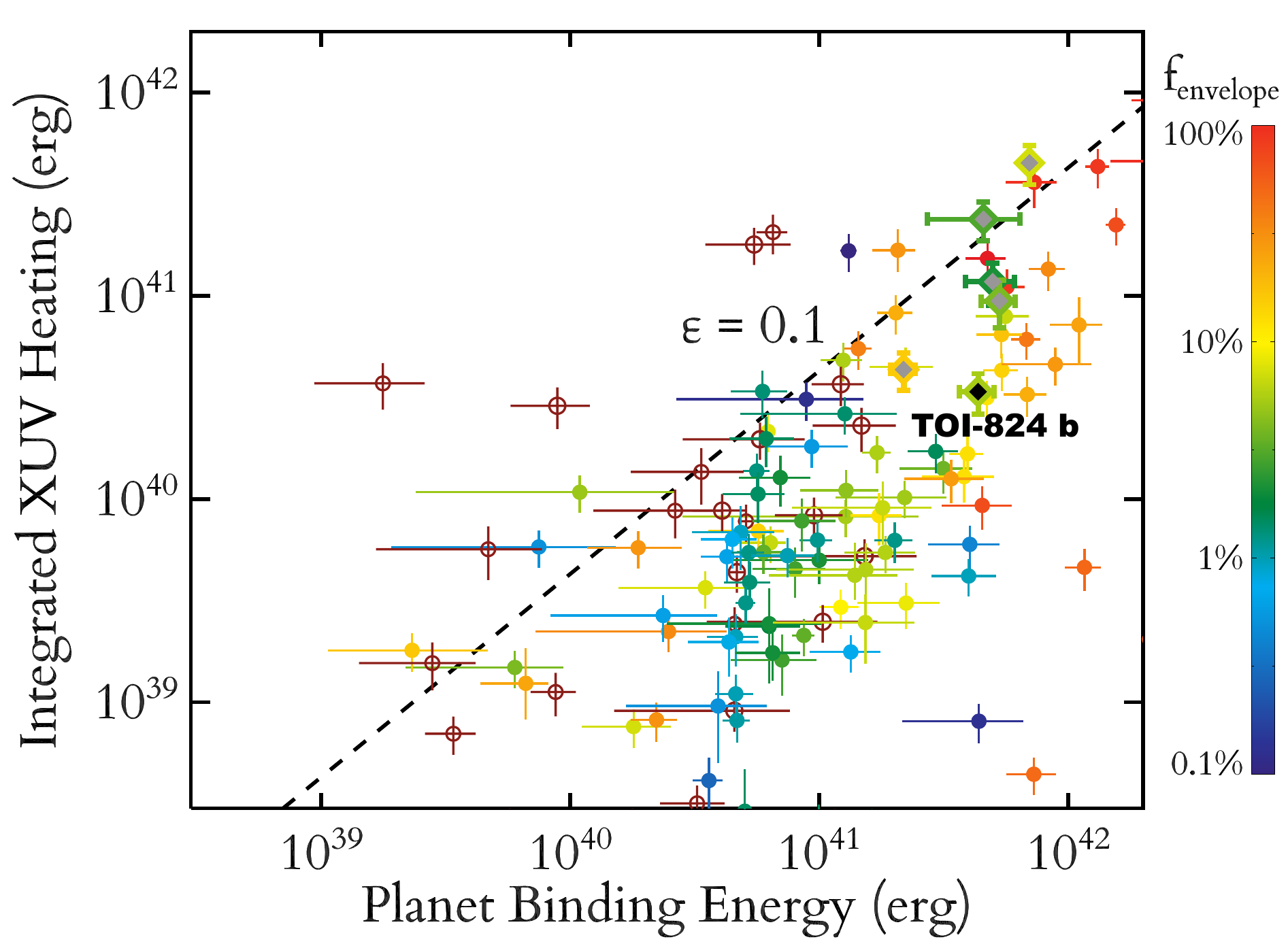}
\caption{Despite their extremely high irradiations, the recent crop of ultra hot Neptunes found in and near the “hot Neptune desert” fit into the wider population of known exoplanets when we consider their exceptionally high masses. Updated from \citet{Lopez2014}, this diagram compares the total XUV heating a planet's atmosphere has received over its lifetime to its gravitational binding energy, with planets at the top left being the most vulnerable to extreme atmospheric escape from photo-evaporation. Planets are color coded based on their estimated H+He gas fraction, with likely bare rocky planets shown by open rust-colored points. Previously known planets are depicted as circles, while the six new high mass hot Neptunes are plotted as diamonds. The dashed line shows a widely used analytic approximation of atmospheric survivability from \citet{Lopez2012} assuming a 10\%\, XUV heating efficiency.
\label{fig:BindingEnergy}}
\vspace{0.7cm}
\end{figure}

Planet evolution and escape models may be able to explain this conundrum. Along with other recent discoveries in and around the desert such as K2-100b \citep{Barragan2019}, HD 219666 b \citep{Esposito2019}, \object{NGTS-4 b} \citep{West2019}, \object{TOI-132 b} \citep{Diaz2019}, and \object{LTT 9779 b} (J. Jenkins, priv. comm.), \object{TOI-824 b} is exceptionally massive given its radius. All of these planets have masses in excess of $16$ \mearth\, despite that fact that planets in this size range (3-5 \rearth), are more typically 6 to 10 \Mearth\, \citep{Wolfgang2016,Ning2018}. Such high masses mean that these planets are more resilient to atmospheric escape since a planet's timescale to lose its atmosphere to photo-evaporative escape scales roughly as M$_{p}^{-2}$ \citep{Lopez2013}. Indeed when viewed in the context of their gravitational binding energy and their XUV irradiation (Figure \ref{fig:BindingEnergy}) these new discoveries appear more typical lying close to but not beyond the limits of potential survival to escape, similar to other previously known hot Neptunes and sub-Neptunes.

Although the large planet mass may help explain how TOI-824 b's atmosphere survived, the existence of these large planet masses alone poses intriguing questions for theorists. Structure models indicate that most of this large mass is likely in the planet's heavy element core \citep{Lopez2014}. Given its extremely short orbital period, however, we must ask how \object{TOI-824 b} and similar planets accumulated such a large amount of heavy elements on such an irradiated orbit in the first place. Theorists have long argued that hot Jupiters likely migrate in from much more distant orbits, however it has been debated whether this is also true of lower mass planets. Studies of ultra short period rocky planets and of the overall distribution of \Kepler\ planets indicate that there is likely some mass enhancement in the inner parts of planetary disks compared to the classic minimum mass solar nebula \citep[e.g.,][]{Chiang2013}. However, with $\sim18.6$ \Mearth\, and an orbital period of only $1.4$ days, systems like \object{TOI-824} may require an even stronger concentration or migration of heavy elements in the inner part of the planetary disk.

\subsection{Potential for atmospheric characterization}

Hot Neptunes are particularly compelling targets for follow-up atmosphere characterization. Their high equilibrium temperatures make it more likely that their atmospheres are cloud free \citep{CrossfieldKreidberg2017}. Their elevated temperatures also mean that they're good targets for thermal emission measurements taken during secondary eclipse, which are less affected by clouds and hazes than transmission spectra \citet{Fortney2005}. The number of Neptune-sized planets in this desired insolation range is currently very limited, however, and only a small number have been studied in depth and had their atmospheres confirmed. Most notable among this population are \object{GJ 436 b} \citep{Butler2004, Morley2017}, \object{GJ 3470 b} \citep{Bonfils2012, Benneke2019}, and \object{HAT-P-11b} \citep{Bakos2010, Fraine2014} which are denoted by the square points in Figure \ref{fig:DesertFig}.

\object{TOI-824 b} is also a compelling target because its mass is precisely known.  \citet{Batalha2019} showed that in order to infer the atmospheric properties of an exoplanet, the planet's mass must be measured to at least the 20\% level. Otherwise the widths of the posterior distributions of the atmospheric properties are dominated by the uncertainties in the planet's mass.  

Absorption features from several key molecular species in the atmosphere of TOI-824 b may be detectable with current ground- and space-based facilities. Hubble/WFC3 observations in the near infrared could reveal water features, assuming a cloud-free, $100\times$ solar metallicity atmosphere. Molecular features from water and CO may also be accessible with high resolution ground-based spectrographs such as CRIRES+ at the VLT \citep{Follert2014}. In addition to these molecular species, \object{TOI-824 b} is hot enough that alkali metals may be present in the gas phase in the atmosphere, in contrast to previously characterized small planets \citep{Morley2015}. The ESPRESSO spectrograph on VLT \citep{Pepe2014}, for example, should be able to detect sodium in the atmosphere of \object{TOI 824 b}. Many additional chemical species will be observable with next-generation facilities like the ELTs that have broader wavelength coverage. 

We note that the expected signal-to-noise for atmospheric features for \object{TOI-824 b}'s is not the highest for all sub-Neptunes discovered by TESS. It sits just barely above the cutoff suggested by \citet{Kempton:2018}, at a transmission spectroscopy metric (TSM) of 85 compared to the suggested inclusion criteria of TSM $\geq$ 84, which was designed to yield a statistical sample of planets in this size range that are accessible with a modest amount of JWST time per planet. That threshold S/N assumes the atmospheres are cloud-free, however, which is not necessarily the case \citep{CrossfieldKreidberg2017}. If \object{TOI-824 b} follows the trend noted in \cite{CrossfieldKreidberg2017}, then it may have relatively large spectral features due to its high temperature. Atmosphere characterization is worth pursuing to test this hypothesis.

\object{TOI-824 b} is also a promising target for the detection of atmospheric escape. At the edge of the hot Neptune desert, the planet has likely experienced significant photoevaporation over its lifetime and into the present. Observations of the helium near-IR triplet may reveal atmospheric escape in action and constrain the rate of evaporative mass loss \citep{Spake2018, Salz2018, Ninan2019}, and similar studies could be carried out using observations of H$\alpha$ \citep{Jensen2012, Cauley2017, Jensen2018, Yan2018}, and Ly$\alpha$ \citep{Ehrenreich2015, Bourrier2018}. Conveniently, \object{TOI-824 b} has a K dwarf host star, which is the optimal stellar spectral type to excite neutral helium atoms \citep{Oklopcic2018,Oklopcic2019}.

As of now, the detectability of \object{TOI-824 b}'s atmosphere from both ground and space is promising and could lead to the detailed characterization of the most irradiated, small planet at the edge of the desert that has retained its atmosphere to date.

\startlongtable
\begin{deluxetable*}{lccc}
\tablecaption{Median values and 68\% confidence interval for EXOFASTv2 results on TOI824.\\
Notes from \citet{Eastman2019}: The star's age is calculated using the MIST isochrones. The optimal conjunction time ($T_{0}$)  is the time of conjunction that minimizes the covariance with the planet's period and therefore has the smallest uncertainty. The equilibrium temperature of the planet ($T_{eq}$) is calculated using Equation 1 of \citet{HansenBarman2007} and assumes no albedo and perfect heat redistribution. The tidal circularization timescale ($\tau_{\rm circ}$) is calculated using Equation 3 from \citet{AdamsLaughlin2006} and assumes Q = 10$^{6}$. The 3.6$\mu$m and 4.6$\mu$m secondary occultation depths use a black-body approximation of the stellar flux, $F_{\star}$, at T$_{\rm{eff}}$ and of the planetary flux, $F_{p}$, at T$_{\rm{eq}}$ and are calculated using $\delta_{S,\lambda}= \frac{(R_{p}/R_{\star})^{2}}{(R_{p}/R_{\star})^{2} + (F_{\star}/F_{p})}$. \label{tab:EXOFASTv2}}
\tablehead{\colhead{~~~Parameter} & \colhead{Units} & \multicolumn{1}{c}{Values}}
\startdata
\smallskip\\\multicolumn{2}{l}{EXOFASTv2 Gaussian priors:}&\smallskip\\
~~~~$R_*$\dotfill &Stellar radius (\rsun) \dotfill &$0.719\pm0.033$\\
~~~~$T_{\rm eff}$\dotfill &Effective Temperature (K) \dotfill &$4569\pm114$\\
~~~~$[{\rm Fe/H}]$\dotfill &Metallicity (dex) \dotfill &$-0.12\pm0.080$\\
~~~~$\varpi$\dotfill &Parallax (mas) \dotfill &$15.696178\pm0.04934$\\
~~~~$\rm{A_{v}}$\dotfill &V-band Extinction (mag) \dotfill &$0.025\pm0.06$\\
\smallskip\\\multicolumn{2}{l}{EXOFASTv2 Hard bounds on parameters:}&\smallskip\\
~~~~$\log{g}$\dotfill &Surface gravity (cgs)\dotfill &[3,5]\\
~~~~$Age$\dotfill &Age (Gyr)\dotfill &[0,10]\\
~~~~$T_{\rm eff}$\dotfill &Effective Temperature (K) \dotfill &[4000,8000]\\
~~~~$[{\rm Fe/H}]$\dotfill &Metallicity (dex) \dotfill &[-1,0.5]\\
\smallskip\\\multicolumn{2}{l}{Stellar Parameters:}&\smallskip\\
~~~~$M_*$\dotfill &Mass (\msun)\dotfill &$0.710^{+0.032}_{-0.031}$\\
~~~~$R_*$\dotfill &Radius (\rsun)\dotfill &$0.695\pm0.027$\\
~~~~$L_*$\dotfill &Luminosity (\lsun)\dotfill &$0.195^{+0.028}_{-0.025}$\\
~~~~$\rho_*$\dotfill &Density (cgs)\dotfill &$2.98^{+0.30}_{-0.27}$\\
~~~~$\log{g}$\dotfill &Surface gravity (cgs)\dotfill &$4.605\pm0.026$\\
~~~~$T_{\rm eff}$\dotfill &Effective Temperature (K)\dotfill &$4600^{+110}_{-100}$\\
~~~~$[{\rm Fe/H}]$\dotfill &Metallicity (dex)\dotfill &$-0.092^{+0.076}_{-0.077}$\\
~~~~$[{\rm Fe/H}]_{0}$\dotfill &Initial Metallicity \dotfill &$-0.080^{+0.077}_{-0.079}$\\
~~~~$Age$\dotfill &Age (Gyr)\dotfill &$7.5^{+1.8}_{-2.9}$\\
~~~~$EEP$\dotfill &Equal Evolutionary Phase \dotfill &$333.7^{+7.1}_{-13}$\\
~~~~$\dot{\gamma}$\dotfill &RV slope (m/s/day)\dotfill &$-0.138\pm0.067$\\
~~~~$A_D$\dotfill &TESS dilution from neighboring stars \dotfill &$-0.26^{+0.11}_{-0.13}$\\
\smallskip\\\multicolumn{2}{l}{Planetary Parameters:}&b\smallskip\\
~~~~$P$\dotfill &Period (days)\dotfill &$1.392978^{+0.000018}_{-0.000017}$\\
~~~~$R_P$\dotfill &Radius (\re)\dotfill &$2.926^{+0.202}_{-0.191}$\\
~~~~$M_P$\dotfill &Mass (\me)\dotfill &$18.467^{+1.843}_{-1.875}$\\
~~~~$T_C$\dotfill &Time of conjunction (\bjdtdb)\dotfill &$2458597.81419^{+0.00063}_{-0.00065}$\\
~~~~$T_0$\dotfill &Optimal conjunction Time (\bjdtdb)\dotfill &$2458639.60354\pm0.00035$\\
~~~~$a$\dotfill &Semi-major axis (AU)\dotfill &$0.02177\pm0.00032$\\
~~~~$i$\dotfill &Inclination (Degrees)\dotfill &$83.65^{+0.39}_{-0.38}$\\
~~~~$T_{eq}$\dotfill &Equilibrium temperature (K)\dotfill &$1253^{+38}_{-37}$\\
~~~~$\tau_{\rm circ}$\dotfill &Tidal circularization timescale (Gyr)\dotfill &$0.57^{+0.23}_{-0.16}$\\
~~~~$K$\dotfill &RV semi-amplitude (m/s)\dotfill &$13.2^{+1.2}_{-1.3}$\\
~~~~$\log{K}$\dotfill &Log of RV semi-amplitude \dotfill &$1.121^{+0.039}_{-0.044}$\\
~~~~$R_P/R_*$\dotfill &Radius of planet in stellar radii \dotfill &$0.0387^{+0.0018}_{-0.0019}$\\
~~~~$a/R_*$\dotfill &Semi-major axis in stellar radii \dotfill &$6.73^{+0.22}_{-0.21}$\\
~~~~$\delta$\dotfill &Transit depth (fraction)\dotfill &$0.00149\pm0.00014$\\
~~~~$Depth$\dotfill &Flux decrement at mid transit \dotfill &$0.00149\pm0.00014$\\
~~~~$\tau$\dotfill &Ingress/egress transit duration (days)\dotfill &$0.00386^{+0.00037}_{-0.00034}$\\
~~~~$T_{14}$\dotfill &Total transit duration (days)\dotfill &$0.04806^{+0.00081}_{-0.00080}$\\
~~~~$T_{FWHM}$\dotfill &FWHM transit duration (days)\dotfill &$0.04419^{+0.00084}_{-0.00085}$\\
~~~~$b$\dotfill &Transit Impact parameter \dotfill &$0.745^{+0.022}_{-0.024}$\\
~~~~$\delta_{S,3.6\mu m}$\dotfill &Blackbody eclipse depth at 3.6$\mu$m (ppm)\dotfill &$86^{+12}_{-11}$\\
~~~~$\delta_{S,4.5\mu m}$\dotfill &Blackbody eclipse depth at 4.5$\mu$m (ppm)\dotfill &$126^{+16}_{-15}$\\
~~~~$\rho_P$\dotfill &Density (cgs)\dotfill &$4.03^{+0.98}_{-0.78}$\\
~~~~$logg_P$\dotfill &Surface gravity \dotfill &$3.323^{+0.069}_{-0.070}$\\
~~~~$\Theta$\dotfill &Safronov Number \dotfill &$0.0136\pm0.0016$\\
~~~~$\fave$\dotfill &Incident Flux (\fluxcgs)\dotfill &$0.561^{+0.070}_{-0.063}$\\
~~~~$T_P$\dotfill &Time of Periastron (\bjdtdb)\dotfill &$2458597.81419^{+0.00063}_{-0.00065}$\\
~~~~$M_P/M_*$\dotfill &Mass ratio \dotfill &$0.0000782^{+0.0000074}_{-0.0000076}$\\
~~~~$d/R_*$\dotfill &Separation at mid transit \dotfill &$6.73^{+0.22}_{-0.21}$\\
\smallskip\\\multicolumn{2}{l}{Telescope Parameters:}&HARPS&PFS\smallskip\\
~~~~$\gamma_{\rm rel}$\dotfill &Relative RV Offset (m/s)\dotfill &$12.9\pm2.4$&$1.3\pm1.0$\\
~~~~$\sigma_J$\dotfill &RV Jitter (m/s)\dotfill &$4.3^{+3.0}_{-2.1}$&$3.21^{+1.3}_{-0.81}$\\
~~~~$\sigma_J^2$\dotfill &RV Jitter Variance \dotfill &$18^{+35}_{-14}$&$10.3^{+9.6}_{-4.5}$\\
~~~~$\rm{RMS}$\dotfill &RMS of RV residuals (m/s)\dotfill &$2.58$&$2.71$\\
\enddata
\end{deluxetable*}

\acknowledgements
This paper includes data collected by the \TESS\ mission. Funding for the \TESS\ mission is provided by the NASA Explorer Program. We acknowledge the use of public \TESS\ Alert data from pipelines at the \TESS\ Science Office and the \TESS\ Science Operations Center. This paper includes data gathered with the 6.5 meter Magellan Telescopes located at Las Campanas Observatory, Chile. This work makes use of observations from the LCOGT network. This study is based on observations collected at the European Southern Observatory under ESO programmes 0103.C-0874 (PI Nielsen) and 0103.C-0449(A). This work has made use of data from the European Space Agency (ESA) mission Gaia (https://www.cosmos.esa.int/gaia), processed by the Gaia Data Processing and Analysis Consortium (DPAC, https://www.cosmos.esa.int/web/gaia/dpac/consortium). Funding for the DPAC has been provided by national institutions, in particular the institutions participating in the Gaia Multilateral Agreement.

Resources supporting this work were provided by the NASA High-End Computing (HEC) Program through the NASA Advanced Supercomputing (NAS) Division at Ames Research Center for the production of the SPOC data products. Some/all of the data presented in this paper were obtained from the Mikulski Archive for Space Telescopes (MAST). Support for MAST for non-HST data is provided by the NASA Office of Space Science via grant NNX13AC07G and by other grants and contracts. This research has made use of the NASA Exoplanet Archive, which is operated by the California Institute of Technology, under contract with the National Aeronautics and Space Administration under the Exoplanet Exploration Program. This research has made use of NASA's Astrophysics Data System. This research has also made use of the Exoplanet Follow-up Observation Program website, which is operated by the California Institute of Technology, under contract with the National Aeronautics and Space Administration under the Exoplanet Exploration Program. This research made use of Astropy, a community-developed core Python package for Astronomy \citep{Astropy2013}. 

Part of this research was carried out at the Jet Propulsion Laboratory, California Institute of Technology, under a contract with the National Aeronautics and Space Administration (80NM0018D0004).
JAB, CXH and MNG acknowledge support from MIT’s Kavli Institute as Torres postdoctoral fellows.
JH acknowledges the support from the Swiss National Science Foundation under grant 200020\_172746.
JVS is supported by funding from the European Research Council (ERC) under the European Union's Horizon 2020 research and innovation programme (project {\sc Four Aces}; grant agreement No 724427). 
Support for this work was provided by NASA through Hubble Fellowship grant HST-HF2-51399.001 awarded to J.K.T. by the Space Telescope Science Institute, which is operated by the Association of Universities for Research in Astronomy, Inc., for NASA, under contract NAS5-26555. 
TD acknowledges support from MIT’s Kavli Institute as a Kavli postdoctoral fellow.
X.D. acknowledges support from the Branco-Weiss Fellowship as well as from the European Research Council (ERC) under the European Union’s Horizon 2020 research and innovation programme (grant agreement No 851555)
This material is based upon work supported by the National Science Foundation Graduate Research Fellowship Program under Grant No. (DGE-1746045). Any opinions, findings, and conclusions or recommendations expressed in this material are those of the author(s) and do not necessarily reflect the views of the National Science Foundation. 

\facilities{TESS, Magellan:Clay (Planet Finder Spectrograph), ESO:3.6m (HARPS), CTIO:1.5m (CHIRON), VLT:Antu (NACO)} 

\software{EXOFAST \citep[v2][]{Eastman2017,Eastman2019}, Tapir \citep{Jensen:2013}, AstroImageJ \citep{Collins:2017}, SpecMatch-emp \citep{Yee2017}, astropy \citep{Astropy2013}, SPECIES \citep{Soto2018}, isochrones python module \citep{Morton2015}, MOOG \citep{moog, Sneden2012}, PEXO \citep{pexo}, EWComputation (https://github.com/msotov/EWComputation)}

\bibliographystyle{apj}


\end{document}